%% file: g210.tex
\documentclass[nofootinbib,twocolumn,aps,superscriptaddress,preprintnumbers]{myRevtex4}
\pdfoutput=1

\usepackage{amsmath,amssymb,amsfonts} % Typical maths resource packages
\usepackage{graphicx}
\usepackage{color}
\usepackage{xspace}
\usepackage{natbib}   
\usepackage{tabularx}
\usepackage{hyperref}

\definecolor{darkblue1}{rgb}{0,0,.4}
\hypersetup{breaklinks=true, 
            colorlinks=true, 
            linkcolor=darkblue1, 
            menucolor=darkblue1, 
            urlcolor=darkblue1,
            citecolor=darkblue1
}

\newcommand{\figsize}{\columnwidth}
\newcommand{\fighspace}{0.4cm}

\bibstyle{plain}
\input defs

\begin{document}

\preprint{\vbox{\hbox{CERN-OPEN-2010-021, LAL 10-155}}}

\title{\boldmath Reevaluation of the Hadronic Contributions to the Muon $g-2$
        and to \aZ}

\author{M.~Davier}
\affiliation{Laboratoire de l'Acc{\'e}l{\'e}rateur Lin{\'e}aire,
             IN2P3/CNRS, Universit\'e Paris-Sud 11, Orsay, France}
\author{A.~Hoecker}
\affiliation{CERN, CH--1211, Geneva 23, Switzerland}
\author{B.~Malaescu\footnote
{
   Now at CERN, CH--1211, Geneva 23, Switzerland.
}}
\affiliation{Laboratoire de l'Acc{\'e}l{\'e}rateur Lin{\'e}aire,
             IN2P3/CNRS, Universit\'e Paris-Sud 11, Orsay, France}
\author{Z.~Zhang}
\affiliation{Laboratoire de l'Acc{\'e}l{\'e}rateur Lin{\'e}aire,
             IN2P3/CNRS, Universit\'e Paris-Sud 11, Orsay, France}

\date{\today}

\begin{abstract}
We reevaluate the hadronic contributions to the muon magnetic anomaly, and 
to the running of the electromagnetic coupling constant at the $Z$-boson mass. We include
new \pp cross-section data from KLOE, all available multi-hadron data from BABAR, 
a reestimation of missing low-energy contributions using results on cross sections 
and process dynamics from BABAR, a reevaluation of all experimental contributions using 
the software package HVPTools together with a reanalysis of inter-experiment
and inter-channel correlations, and a reevaluation of the continuum contributions 
from perturbative QCD at four loops. These improvements lead to a decrease in the 
hadronic contributions with respect to earlier evaluations. For the muon $g-2$ we 
find lowest-order hadronic contributions of $(692.3 \pm 4.2)\cdot10^{-10}$ 
and $(701.5 \pm 4.7)\cdot10^{-10}$ for the \ee-based and $\tau$-based analyses, respectively,
and full Standard Model predictions that differ by $3.6\,\sigma$ and $2.4\,\sigma$
from the experimental value. For the \ee-based five-quark hadronic contribution 
to \aZ we find $\dahadZf=(275.7\pm1.0)\cdot10^{-4}$. The 
reduced electromagnetic coupling strength at $M_Z$ leads to an increase by 
$7\:\gev$ in the central value of the Higgs boson mass obtained by the standard 
Gfitter fit to electroweak precision data.
\end{abstract}

\maketitle

\section{Introduction}

The Standard Model (SM) predictions of the anomalous magnetic moment of the muon, 
\amu, and of the running electromagnetic coupling constant, $\alpha(s)$,  are 
limited in precision by contributions from virtual hadronic vacuum polarisation. 
The dominant hadronic terms can be calculated with a combination of experimental cross 
section data, involving \ee annihilation to hadrons, and perturbative QCD. These are 
used to evaluate an energy-squared dispersion integral, ranging from the $\piz\gamma$ 
threshold to infinity. The integration kernels occurring in the dispersion relations 
emphasise low photon virtualities, owing to the $1/s$ descend of the cross section, 
and, in case of \amu, to an additional $1/s$ suppression. In the latter case, about 73\% 
of the lowest order hadronic contribution is provided by the $\ppg$ final state,\footnote
{
   Throughout this paper, final state photon radiation is implied for all 
   hadronic final states.
} 
while this channel amounts to only 13\% of the hadronic contribution to $\alpha(s)$ 
at $s=\mZ^2$.

In this paper, we reevaluate the lowest-order hadronic contribution, \amuhadLO, to 
the muon magnetic anomaly, and the hadronic contribution, \dahadZ, to the running 
\aZ at the $Z$-boson mass. We include new \pp cross-section data from KLOE~\cite{kloe10} 
and all the available multi-hadron data from 
BABAR~\cite{babarpipi,babar3pi,babar4pi,babar2pi2pi0,babar4pipi0,babar6pi,babarkkpi,babarkkpipi}. 
We also perform a reestimation of missing low-energy contributions using results on 
cross sections and process dynamics from BABAR. We reevaluate all the experimental 
contributions using the software package HVPTools~\cite{g209}, including a 
comprehensive reanalysis of inter-experiment and inter-channel correlations.
Furthermore, we recompute the continuum contributions using perturbative QCD at 
four loops~\cite{chetkuehn}. These improvements taken together 
lead to a decrease of the hadronic contributions with respect to our earlier 
evaluation~\cite{g209}, and thus to an accentuation of the discrepancy between the 
SM prediction of \amu and the experimental result~\cite{bnl}. The 
reduced electromagnetic coupling strength at $M_Z$ leads to an increase in the most 
probable value for the Higgs boson mass returned by the electroweak fit, thus 
relaxing the tension with the exclusion results from the direct Higgs searches. 

\section{New Input Data}

\begin{figure*}[p]
\begin{center}
\includegraphics[width=130mm]{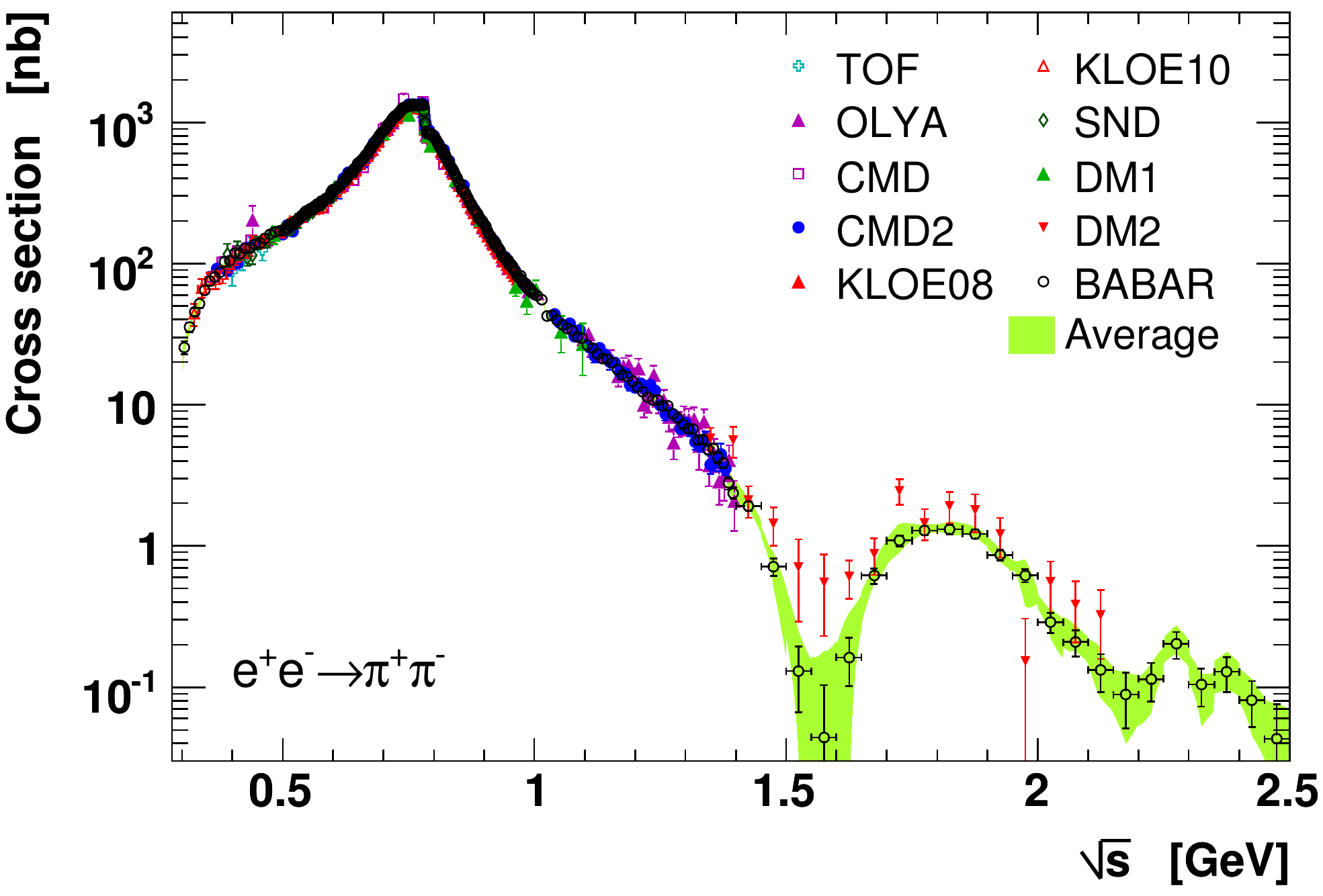}
\vspace{0.5cm}

\includegraphics[width=\figsize]{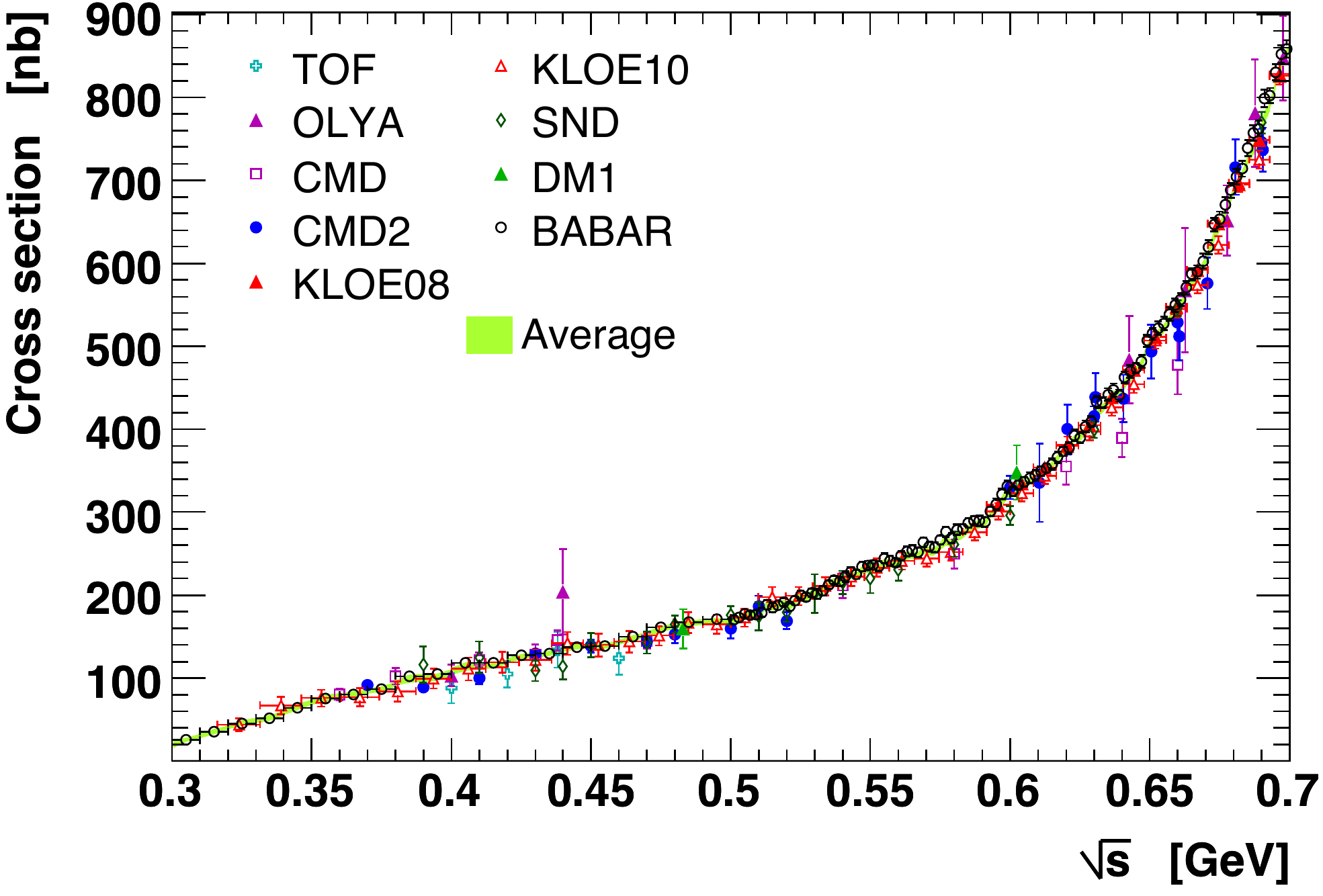}\hspace{\fighspace}
\includegraphics[width=\figsize]{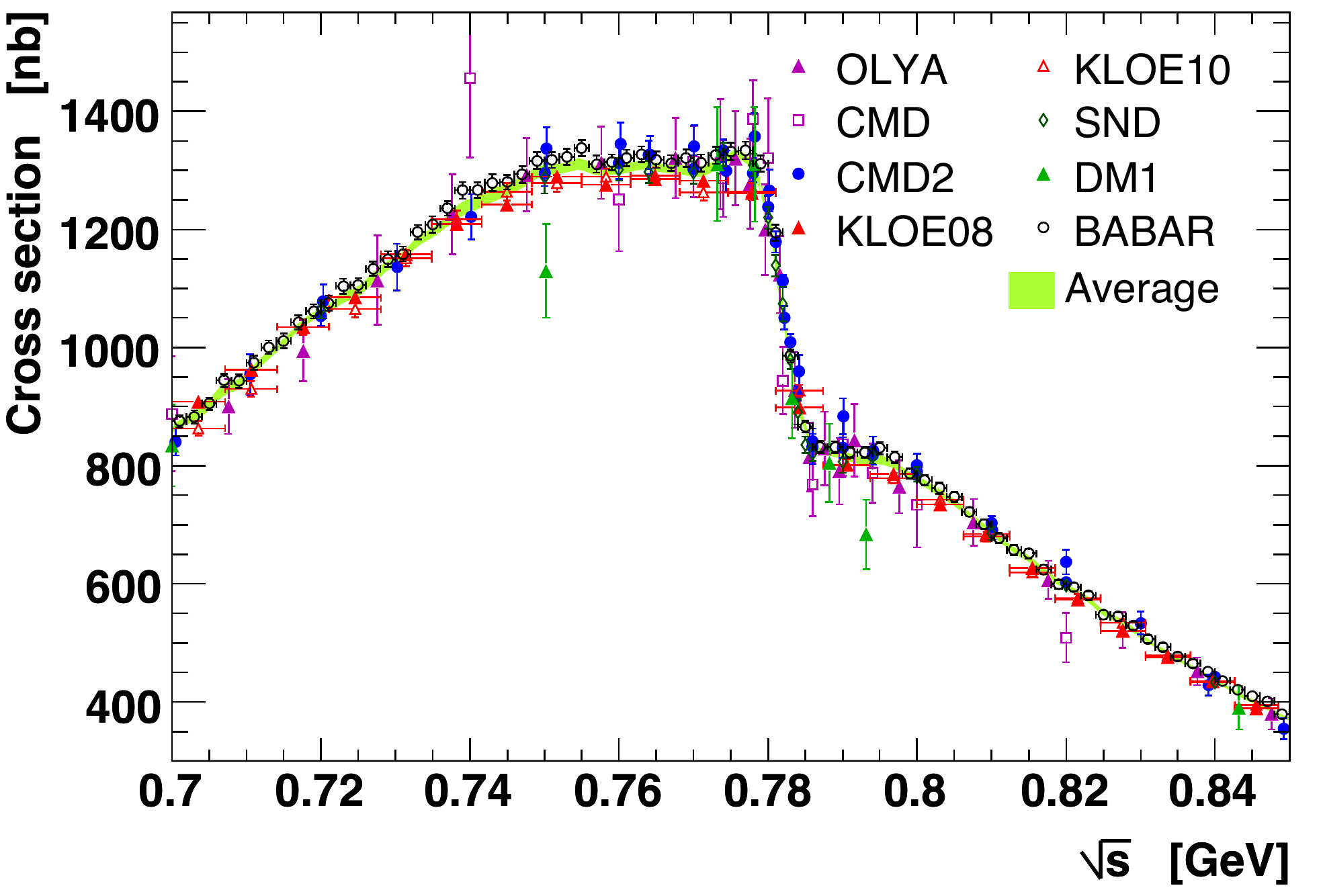}
\vspace{0.2cm}

\includegraphics[width=\figsize]{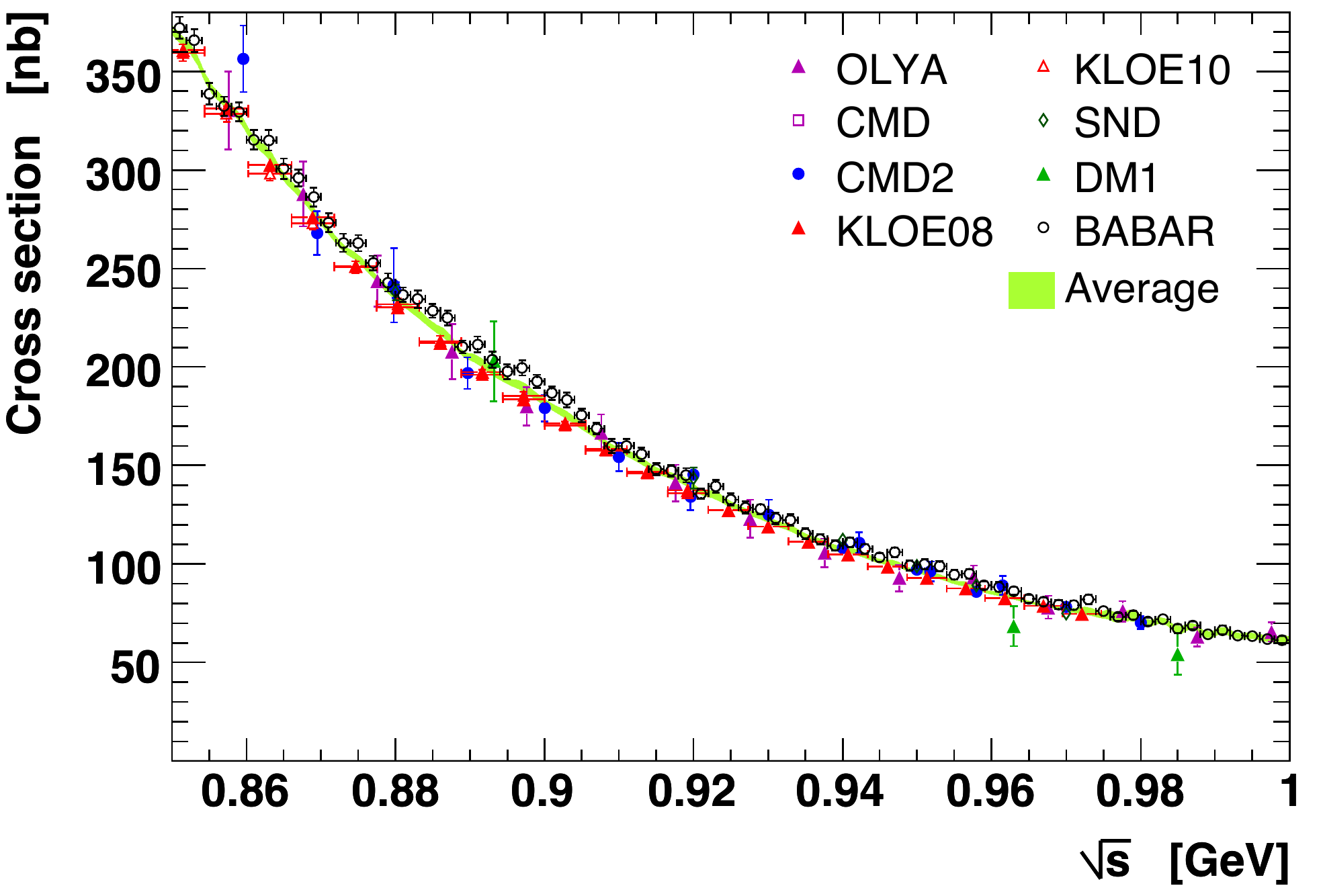}\hspace{\fighspace}
\includegraphics[width=\figsize]{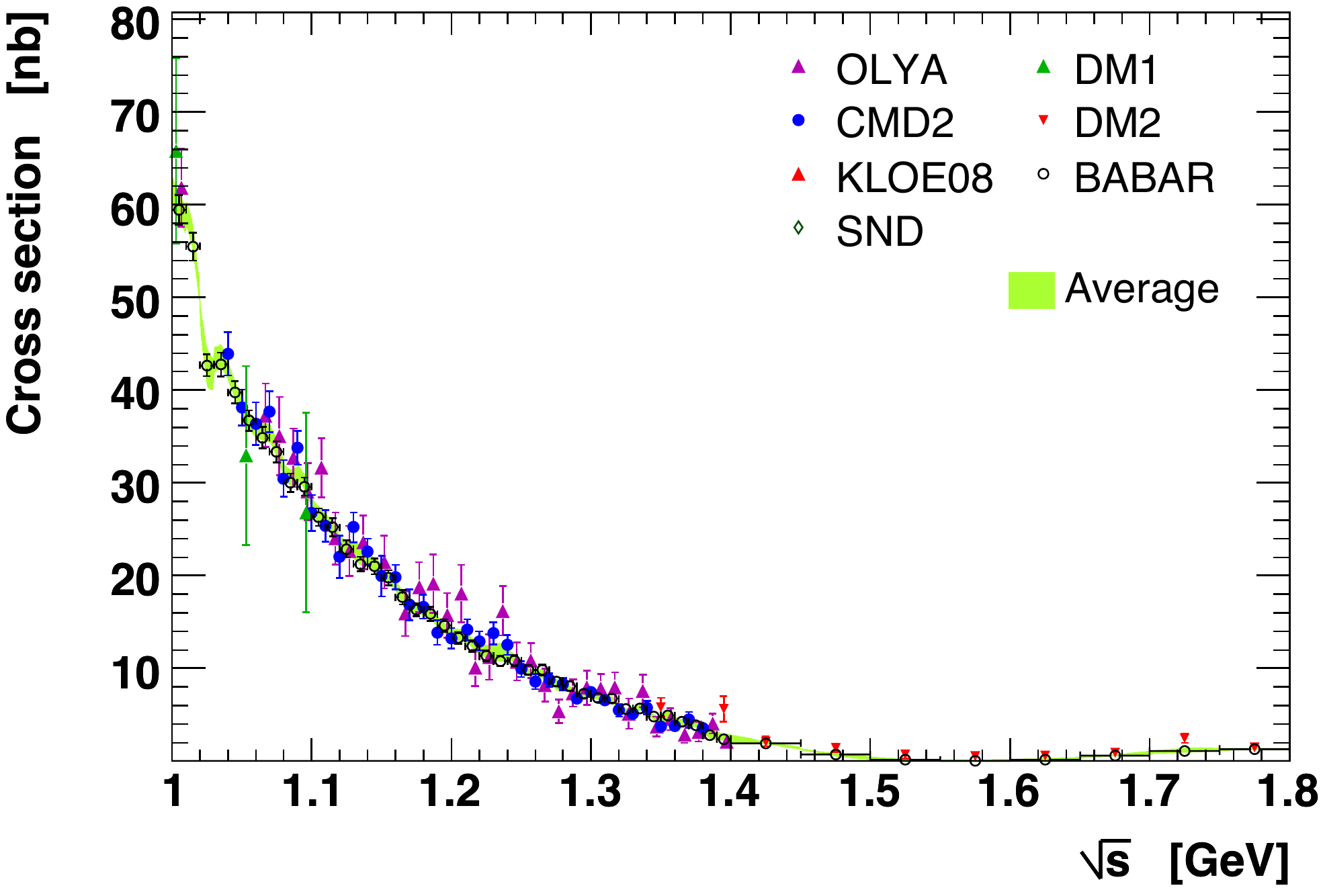}
\end{center}
\vspace{-0.4cm}
\caption[.]{ 
            Cross section of $\ee\to\pp$ versus centre-of-mass energy for different 
            energy ranges. Shown are data from 
            TOF~\cite{E_55}, OLYA~\cite{barkov,E_54}, CMD~\cite{barkov}, 
            CMD2~\cite{cmd203,cmd2new}, SND~\cite{snd}
            DM1~\cite{E_58}, DM2~\cite{E_59}, KLOE~\cite{kloe08,kloe10}, and 
            BABAR~\cite{babarpipi}. The error bars show statistical and systematic 
            errors added in quadrature. The light shaded (green) band indicates
            the HVPTools average within $1\,\sigma$ errors. 
    
}
\label{fig:xsecpipi}
\end{figure*}
\begin{figure*}[t]
\begin{center}
\includegraphics[width=\figsize]{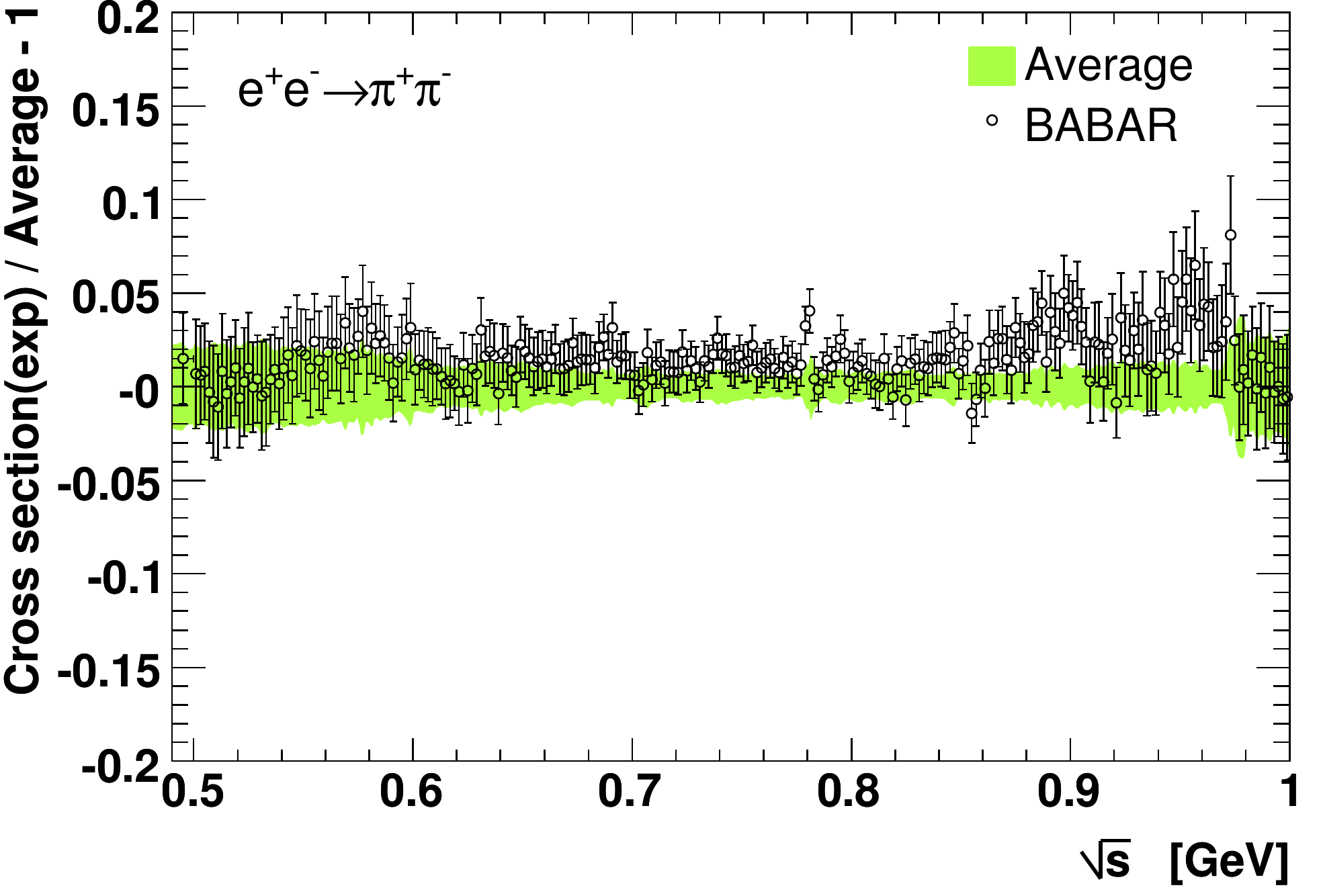}\hspace{\fighspace}
\includegraphics[width=\figsize]{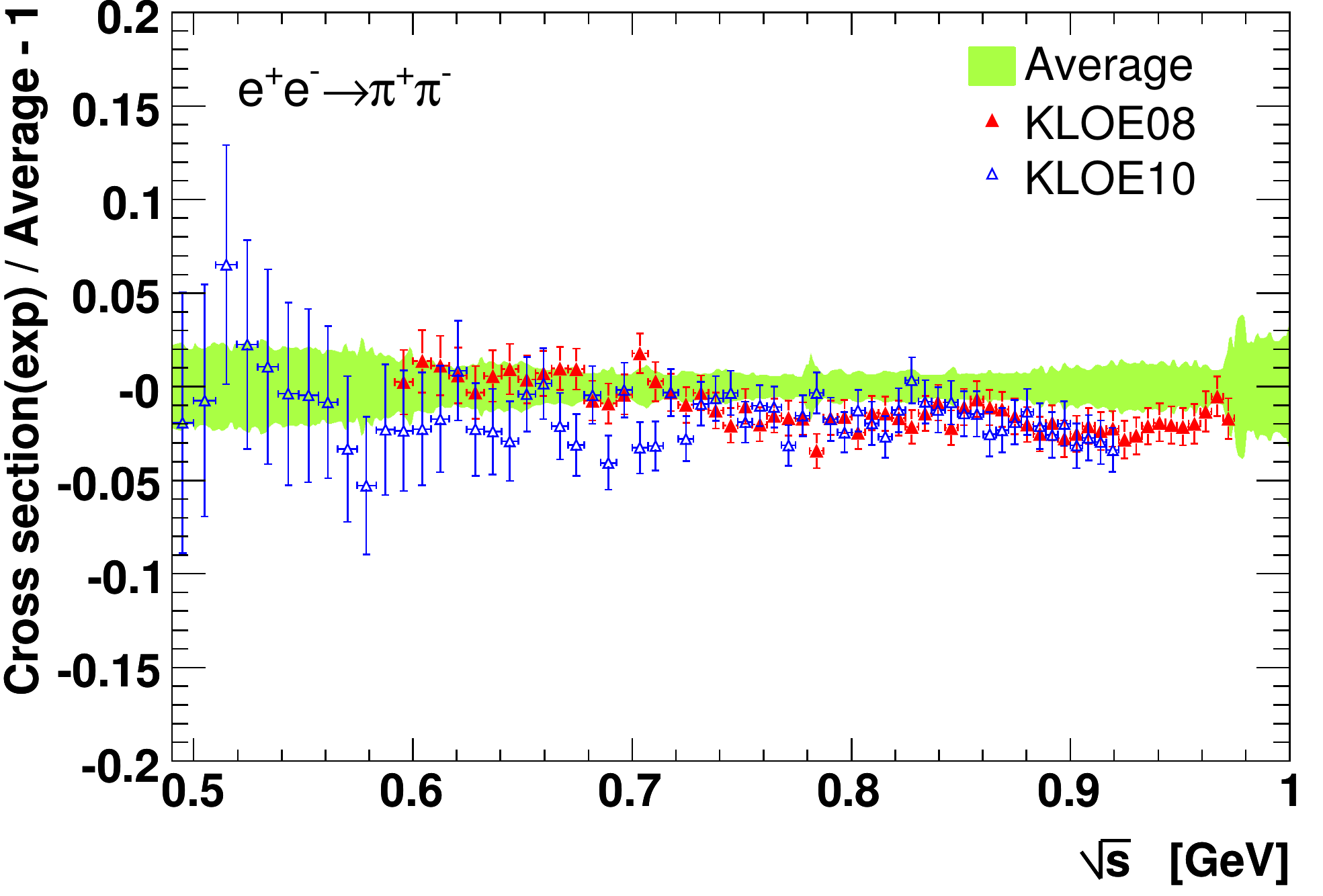}
\vspace{0.2cm}

\includegraphics[width=\figsize]{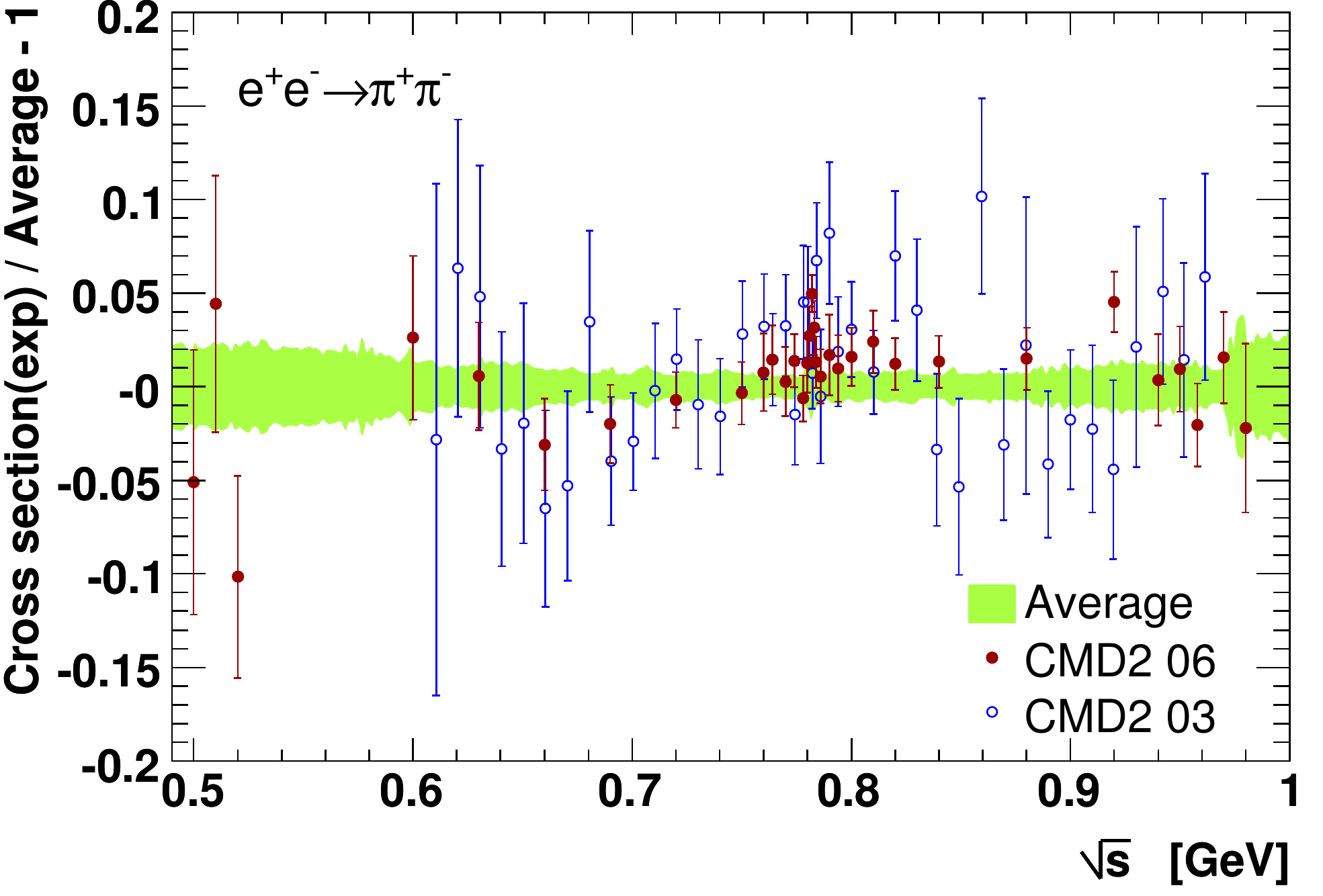}\hspace{\fighspace}
\includegraphics[width=\figsize]{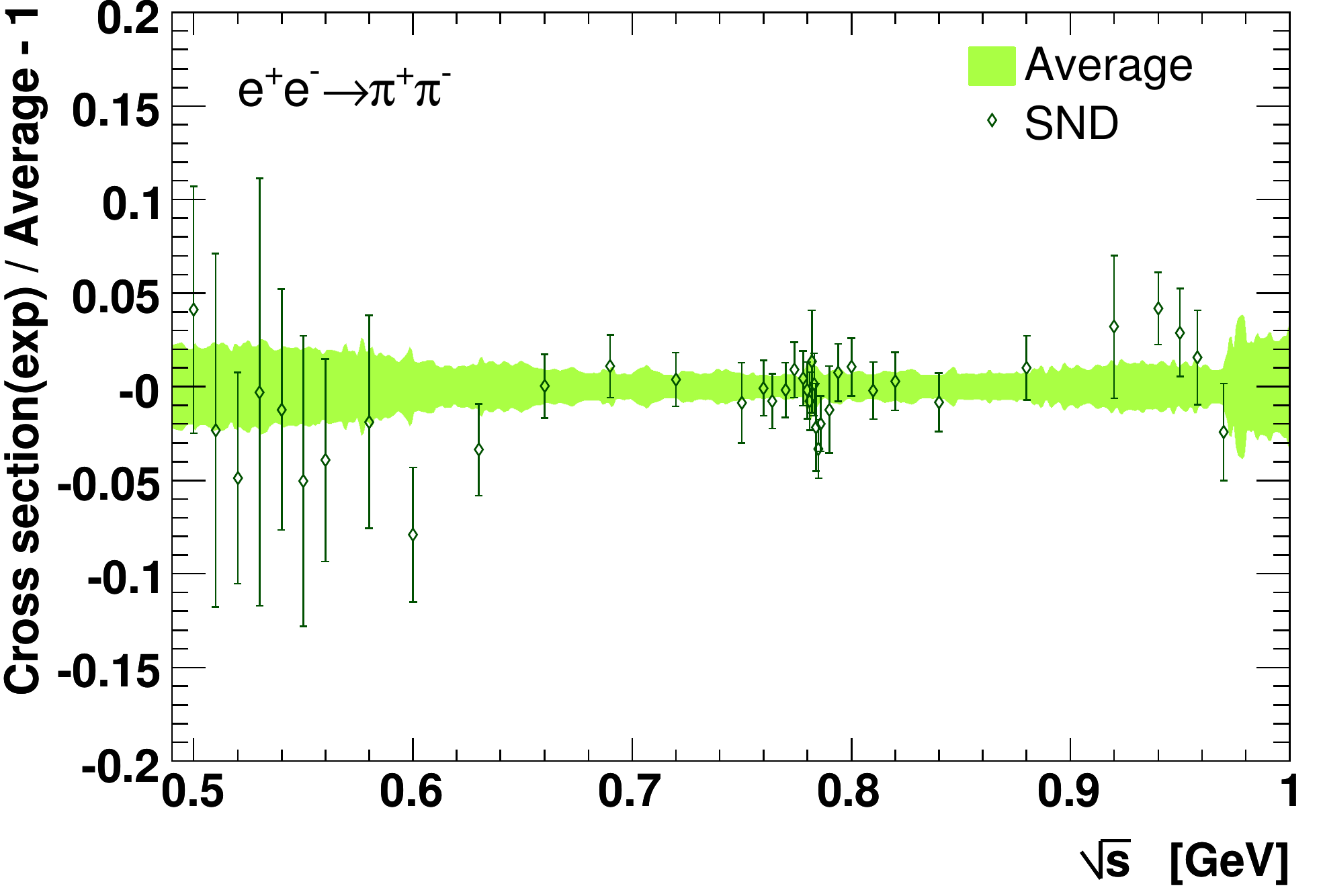}
\end{center}
\vspace{-0.5cm}
\caption[.]{ 
            Comparison between individual $\ee\to\pp$ cross section measurements from 
            BABAR~\cite{babarpipi}, KLOE\,08~\cite{kloe08}, KLOE\,10~\cite{kloe10},
            CMD2\,03~\cite{cmd203}, CMD2\,06~\cite{cmd2new}, SND~\cite{snd}, and 
            the HVPTools average. The error bars show statistical and systematic 
            errors added in quadrature. 
}
\label{fig:comppipi}
\end{figure*}
\begin{figure}[t]
\includegraphics[width=\figsize]{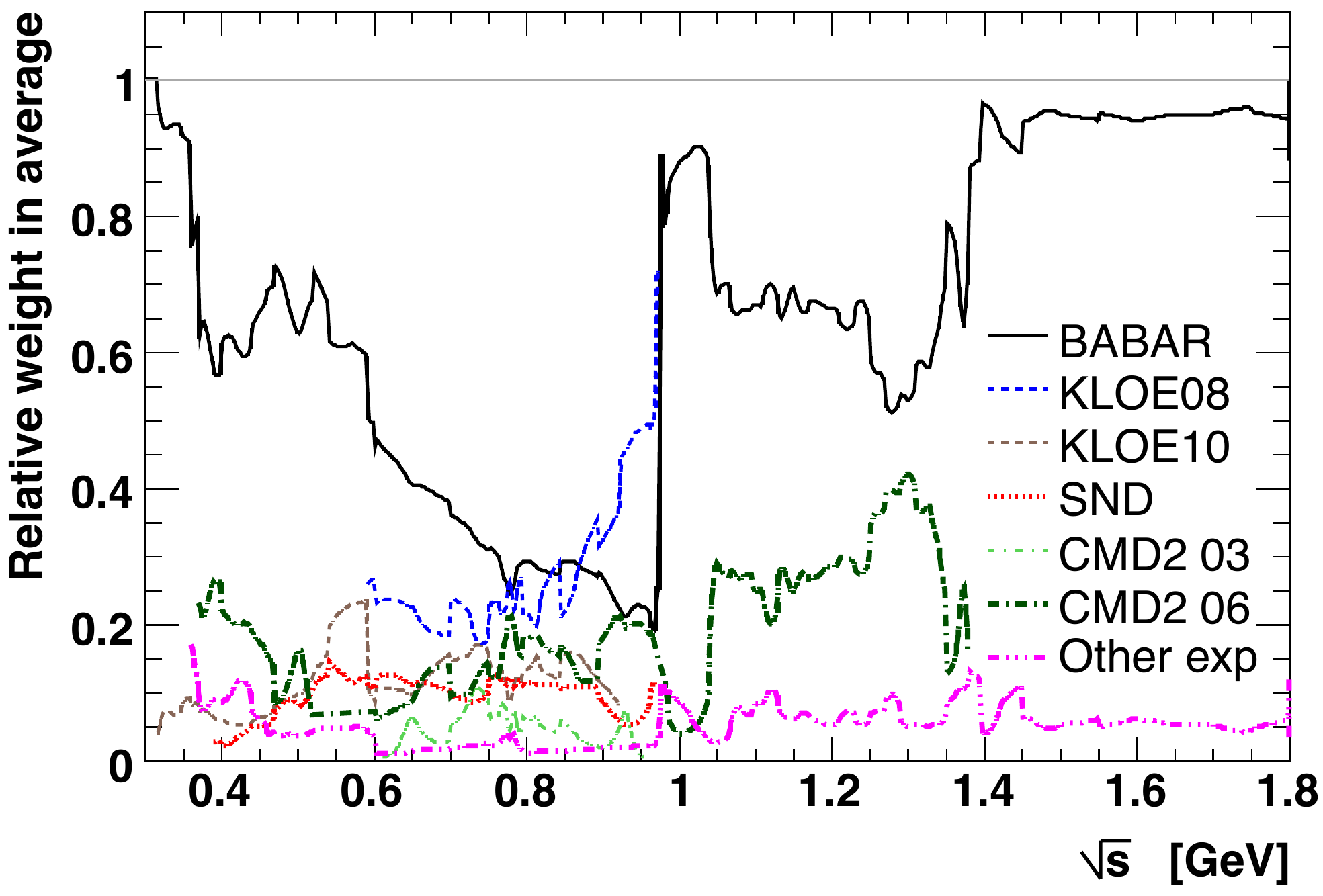}
\vspace{-0.5cm}
\caption[.]{ 
            Relative local averaging weight per experiment versus centre-of-mass energy 
            in $\ee\to\pp$. See Figs.~\ref{fig:xsecpipi} and \ref{fig:comppipi} for references. }
\label{fig:weights}
\end{figure}
\begin{figure}[t]
\begin{center}
\includegraphics[width=\figsize]{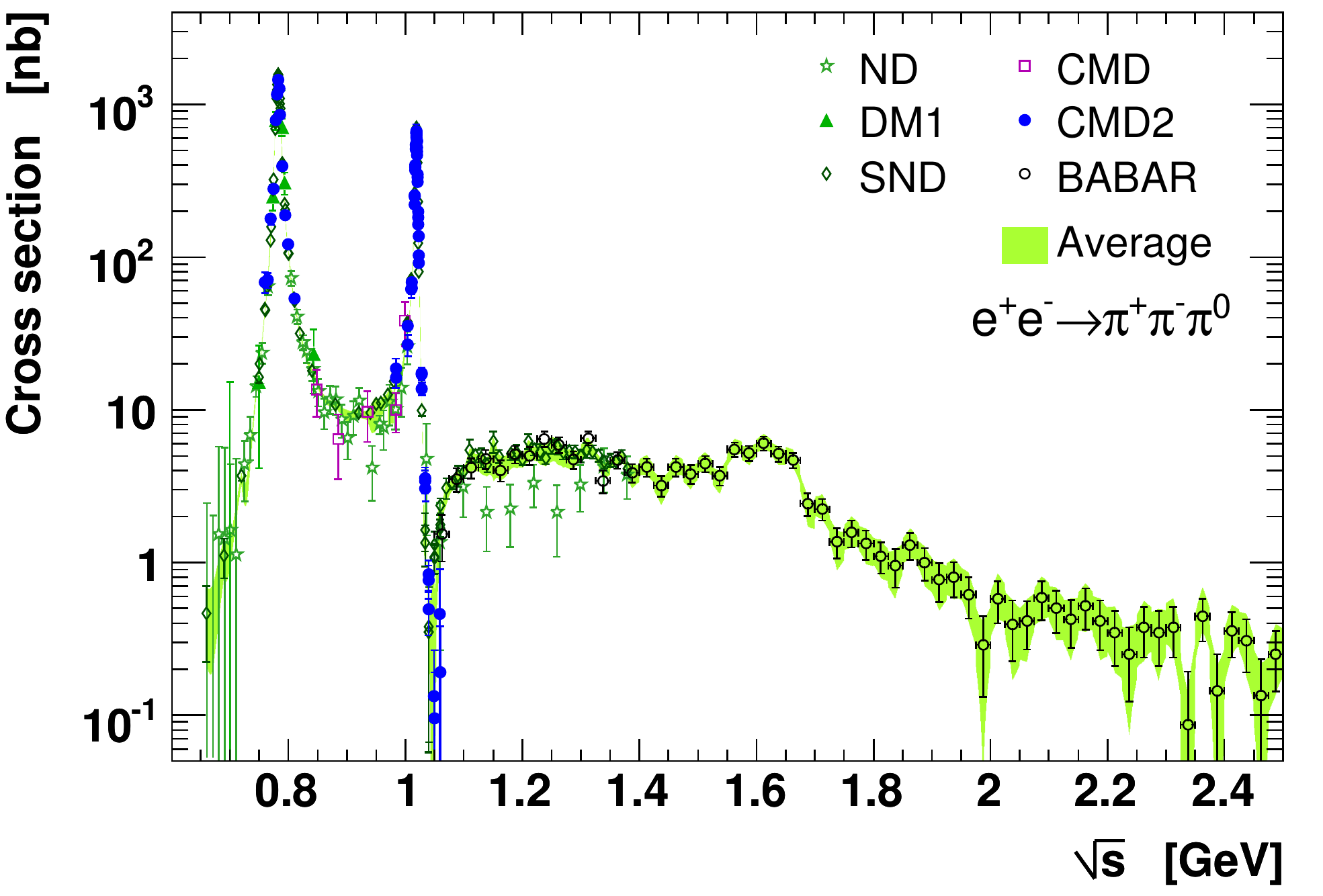}
\end{center}
\vspace{-0.5cm}
\caption[.]{ 
            Cross section of $\ee\to\pp\piz$ versus centre-of-mass energy.
            Shown are data from ND~\cite{nd}, DM1~\cite{dm13pi}, SND~\cite{snd3pi}
            CMD~\cite{cmd3pi}, CMD2~\cite{cmd23pi} and BABAR~\cite{babar3pi}.
            The error bars show statistical and systematic 
            errors added in quadrature. The light shaded (green) band indicates 
            the HVPTools average within $1\,\sigma$ errors.     
}
\label{fig:xsec3pi}
\end{figure}
\begin{figure*}[t]
\begin{center}
\includegraphics[width=\figsize]{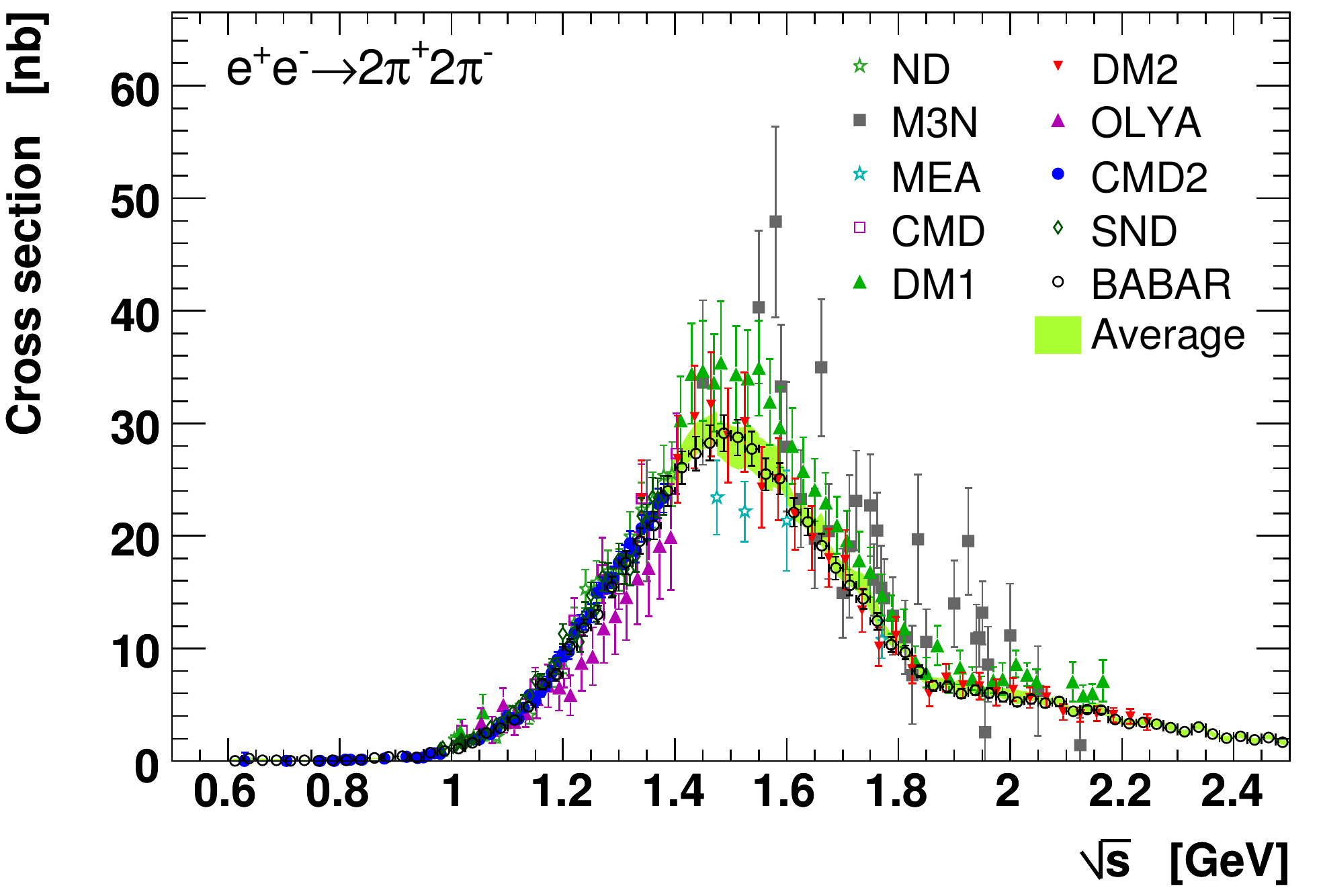}\hspace{\fighspace}
\includegraphics[width=\figsize]{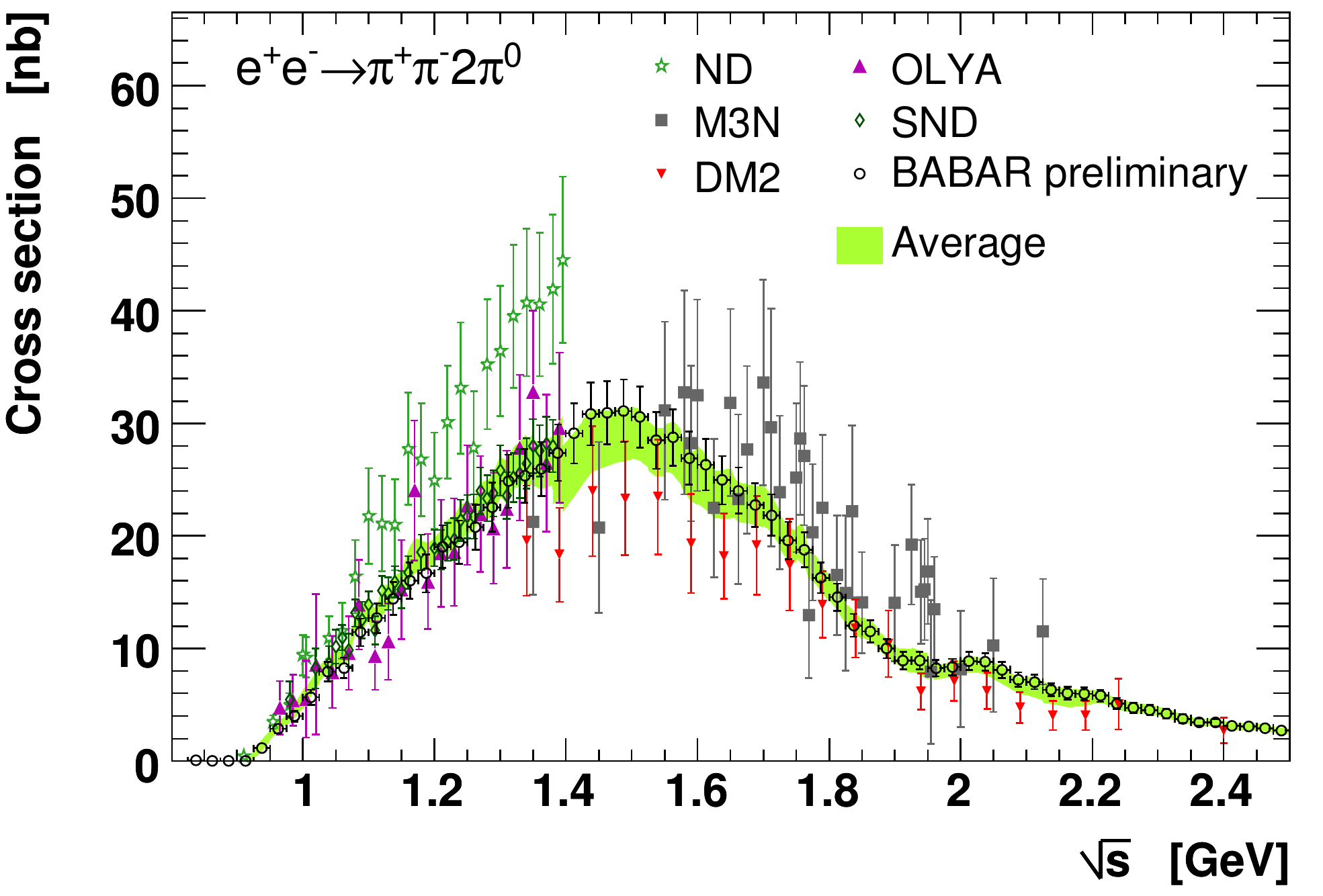}
\vspace{0.2cm}

\includegraphics[width=\figsize]{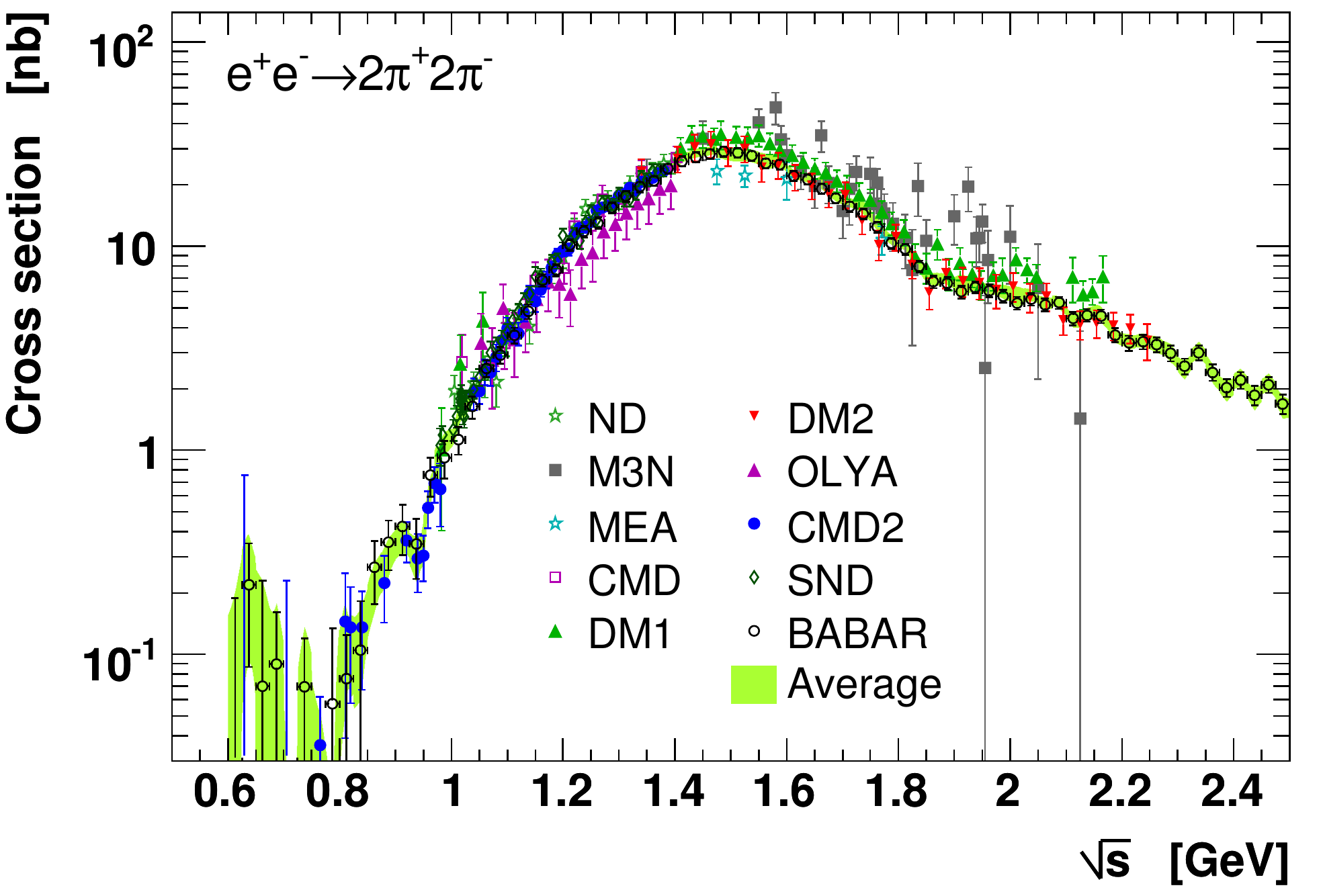}\hspace{\fighspace}
\includegraphics[width=\figsize]{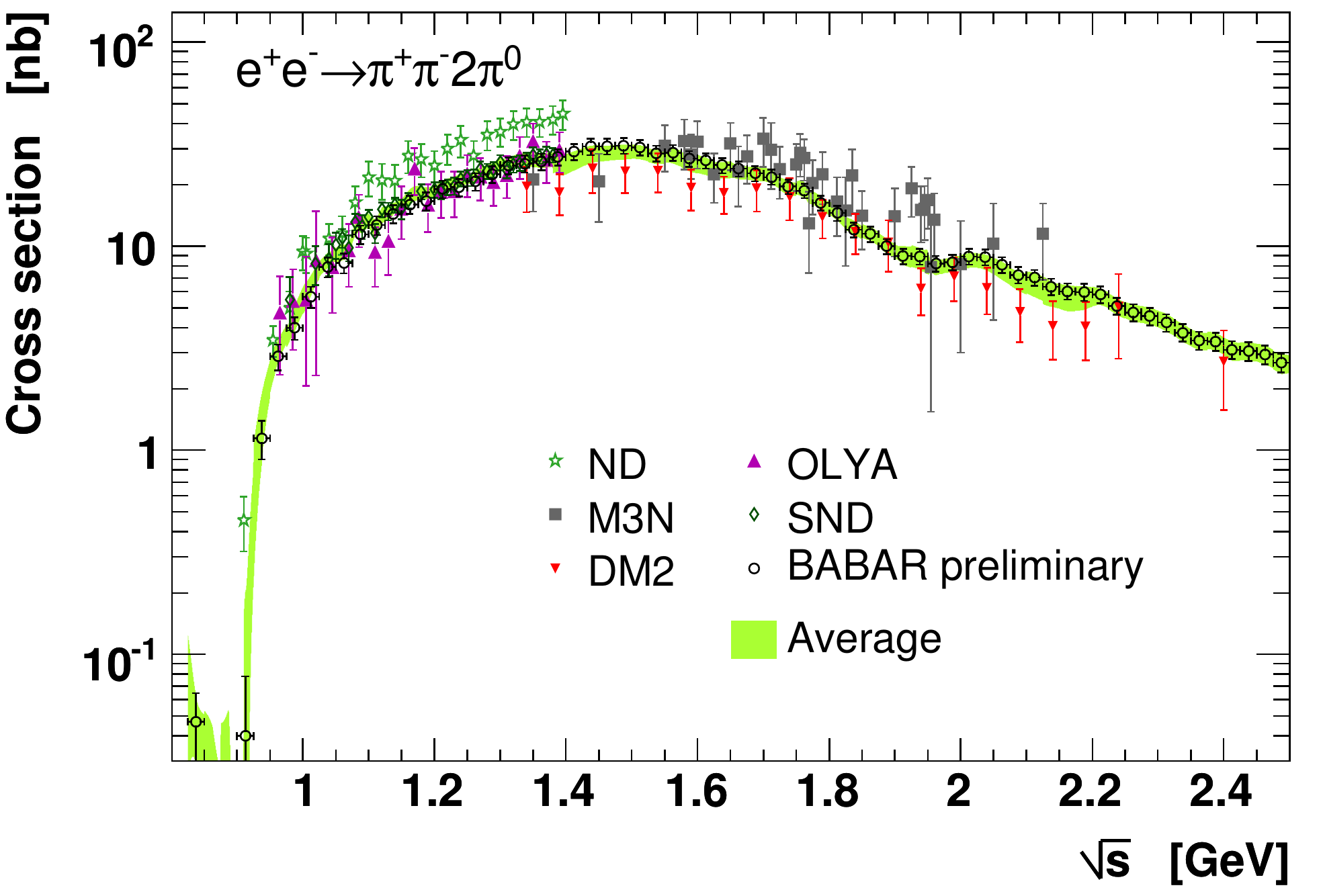}
\end{center}
\vspace{-0.5cm}
\caption[.]{ 
            Cross section versus centre-of-mass energy of $\ee\to 2\pip 2\pim$ (left) and 
            $\ee\to\pp 2\piz$ (right), and for linear (top) and logarithmic 
            ordinates (bottom). The open circles show data from 
            BABAR~\cite{babar4pi,babar2pi2pi0}, which dominate in precision. 
            The other measurements shown are taken for the four charged pions final 
            state from 
            ND~\cite{E_64}, M3N~\cite{m3n4pi}, MEA~\cite{mea4pi}, CMD~\cite{E_72}, 
            DM1~\cite{E_73,PhysLett81B}, DM2~\cite{E_74,E_74p,E_75}, OLYA~\cite{E_70}, 
            CMD2~\cite{cmd2_2pi2pi0} and SND~\cite{snd_2pi2pi0}, and 
            for the mixed charged and neutral state from 
            ND~\cite{E_64}, M3N~\cite{E_66}, DM2~\cite{E_74,E_74p,E_75}, OLYA~\cite{ol}, 
            and SND~\cite{snd_2pi2pi0}. The error bars show statistical and systematic 
            errors added in quadrature. The shaded (green) bands give the HVPTools averages.
}
\label{fig:xsec4pi}
\end{figure*}
\begin{figure*}[t]
\begin{center}
\includegraphics[width=\figsize]{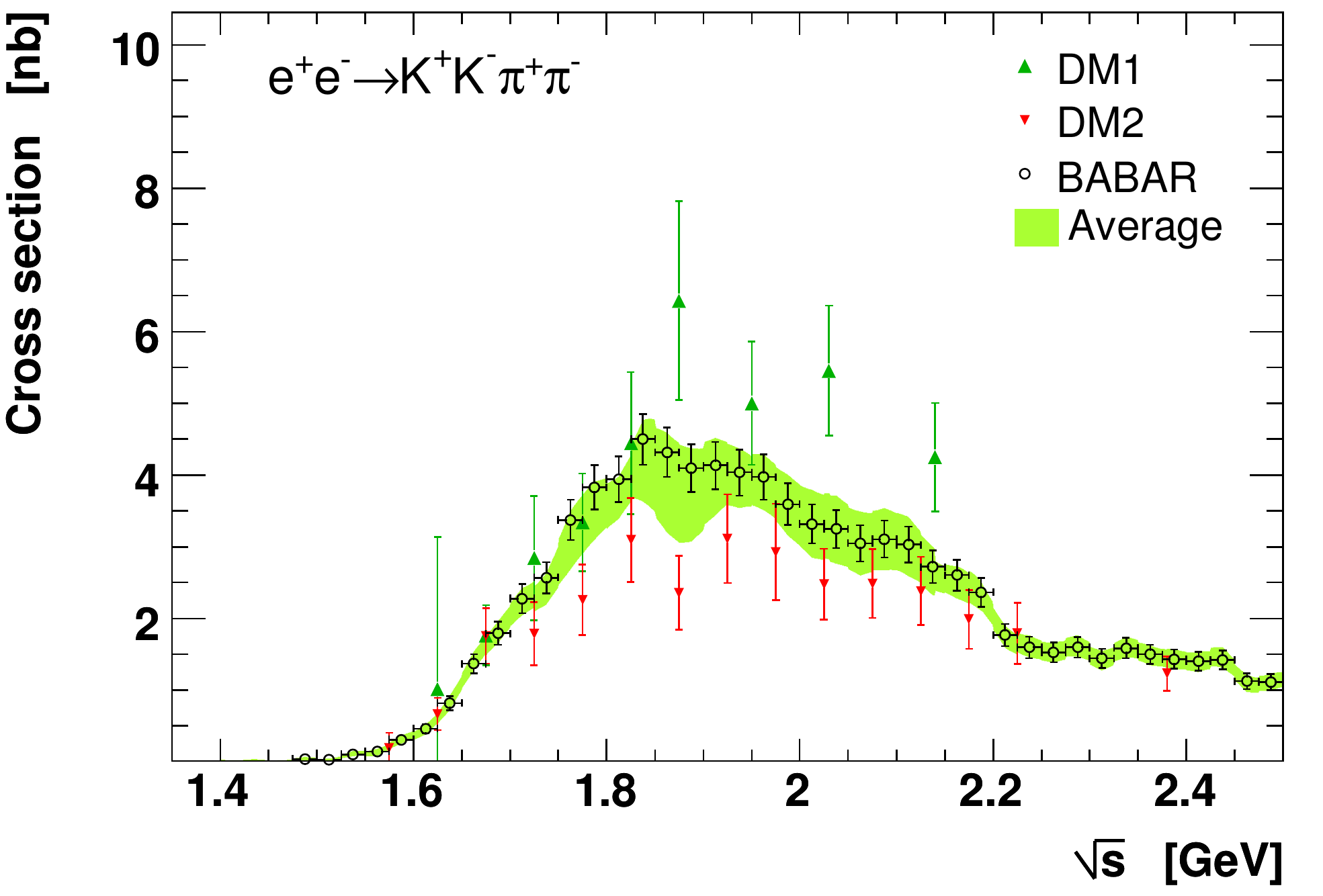}\hspace{\fighspace}
\includegraphics[width=\figsize]{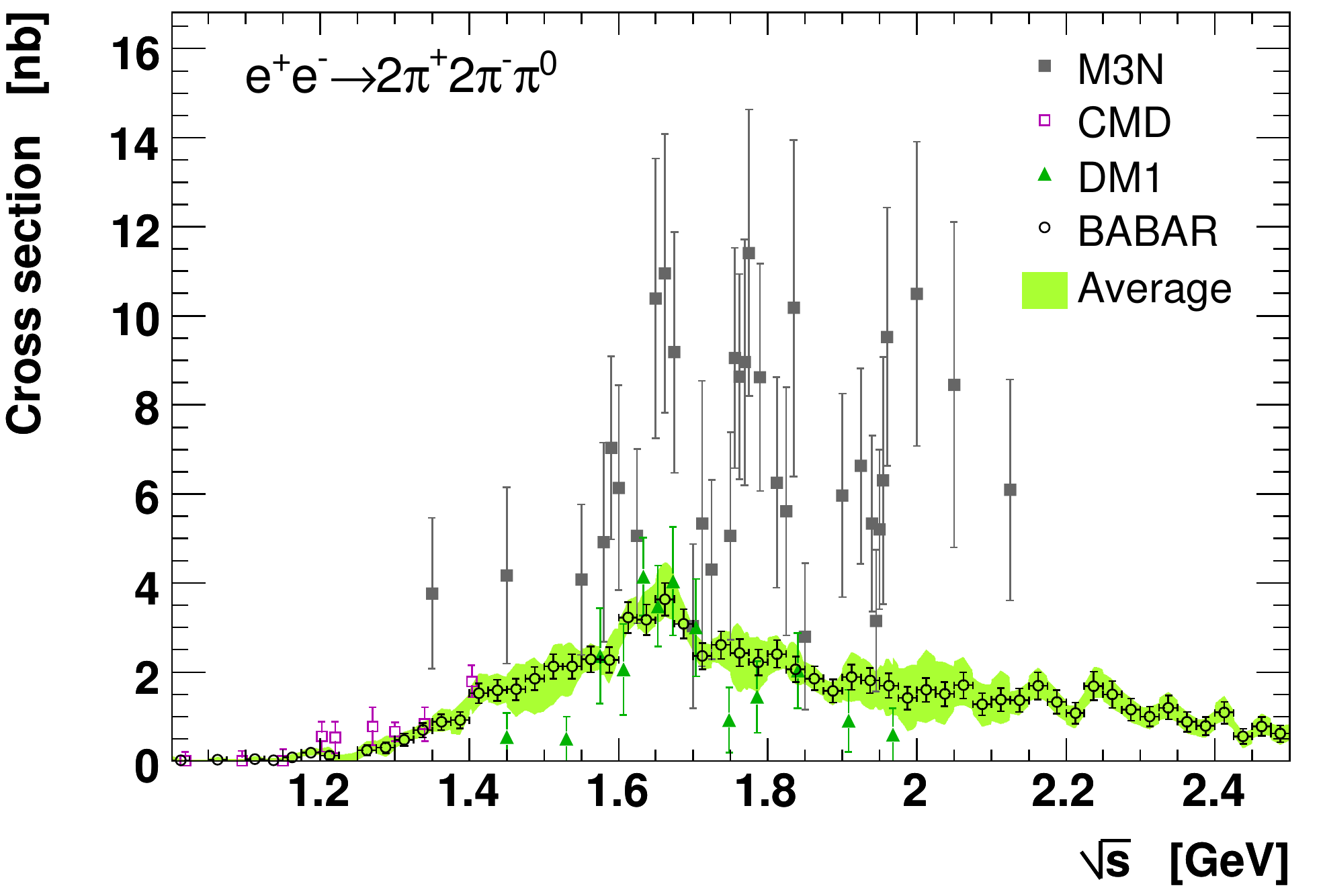}
\vspace{0.2cm}

\includegraphics[width=\figsize]{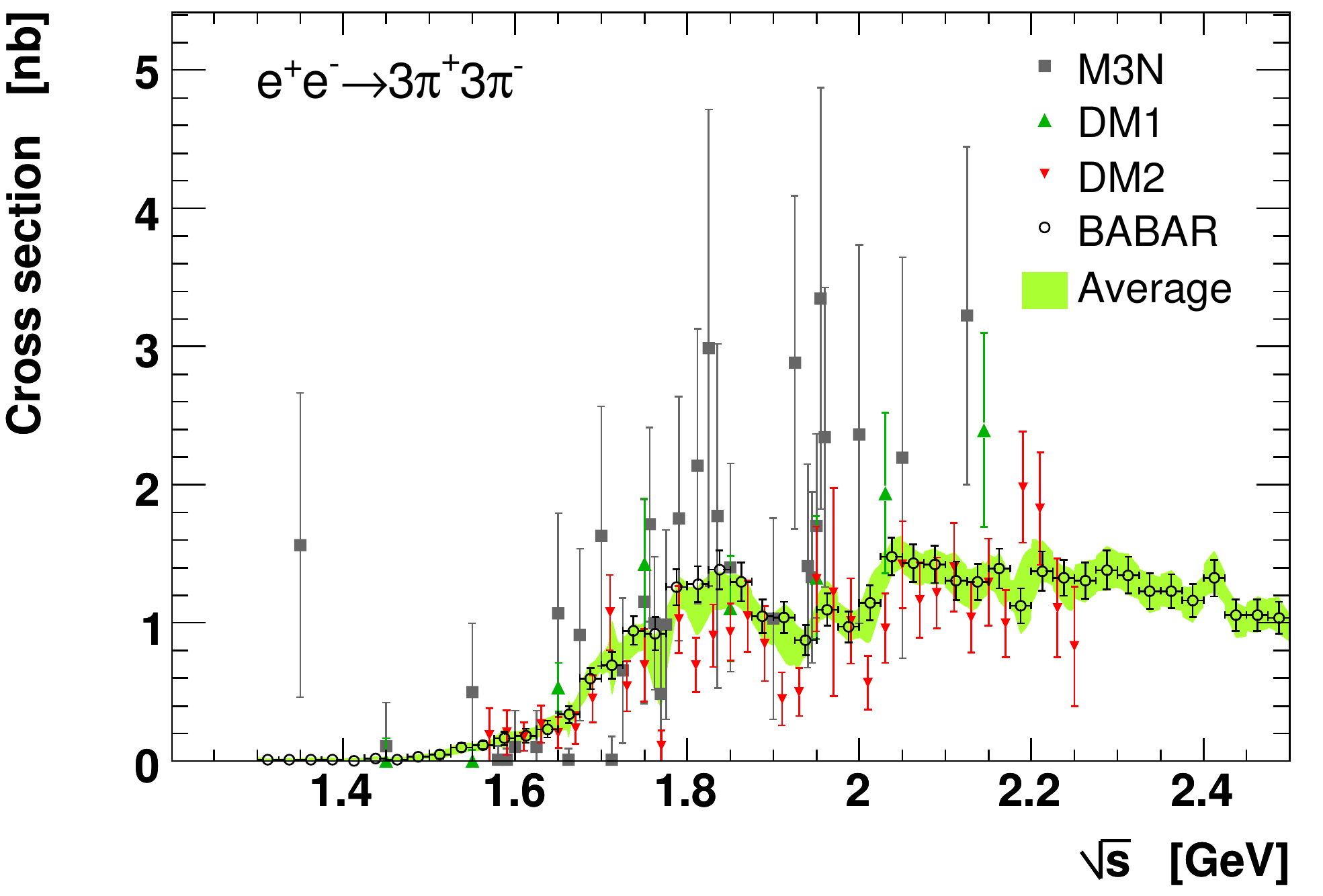}\hspace{\fighspace}
\includegraphics[width=\figsize]{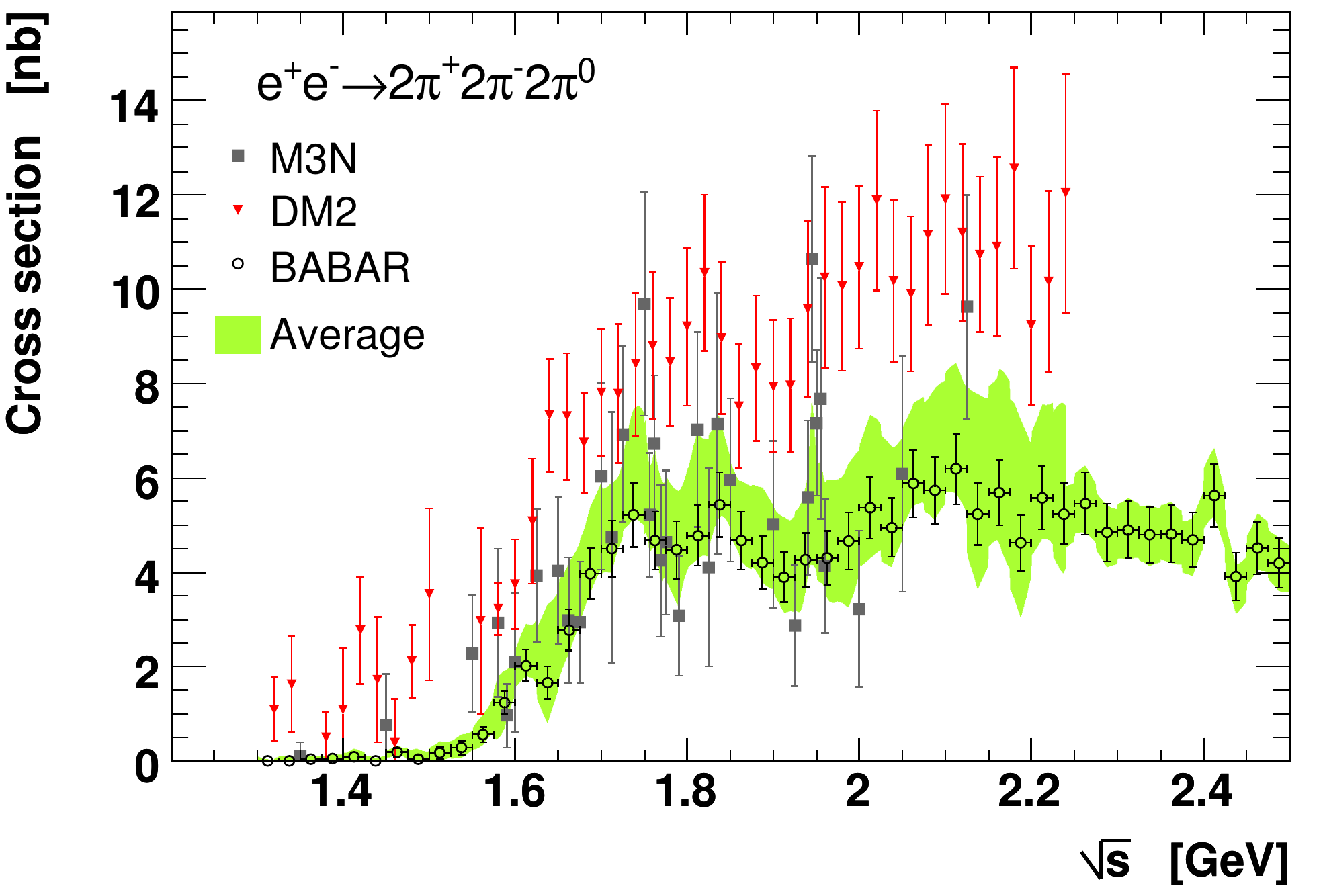}
\end{center}
\vspace{-0.5cm}
\caption[.]{ 
            Cross section data for the final states: $\Kp\Km\pp$ (upper left),
            $2\pip2\pim\piz$ (upper right), $3\pip3\pim$ (lower left) and
            $2\pip2\pim2\piz$ (lower right). 
            The BABAR data points are taken from Refs.~\cite{babar4pi,babar4pipi0,babar6pi}.
            All the other measurements are referenced in~\cite{dehz02,dehz03}.
            The shaded (green) bands give the HVPTools averages within $1\,\sigma$ errors, 
            locally rescaled in case of incompatibilities. 
}
\label{fig:xsecmulti}
\end{figure*}
\begin{figure}[t]
\begin{center}
\includegraphics[width=\figsize]{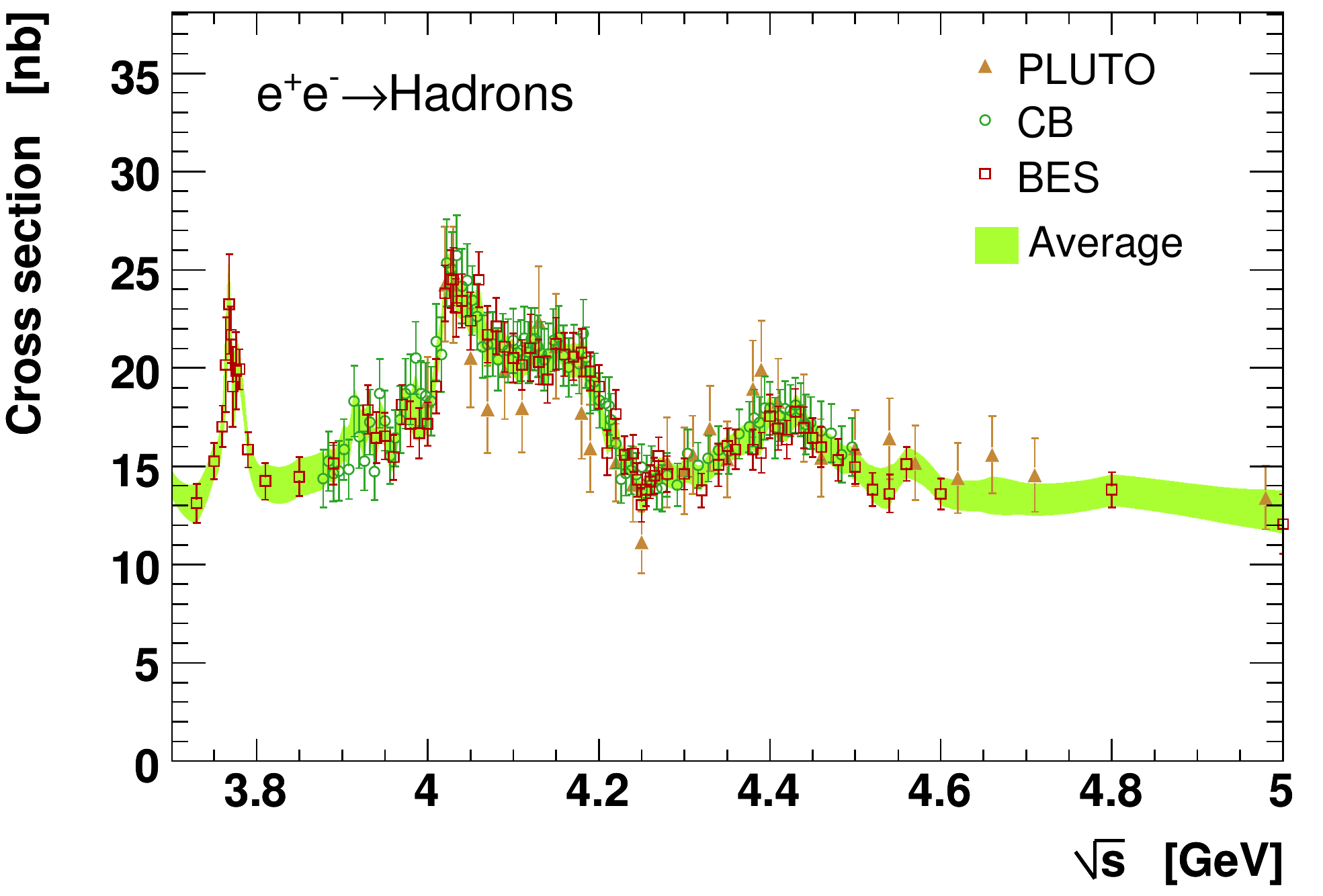}
\end{center}
\vspace{-0.5cm}
\caption[.]{ 
            Inclusive hadronic cross section versus centre-of-mass energy above the 
            $D\Dbar$ threshold. The measurements are taken from 
            PLUTO~\cite{plutoR}, Crystal Ball~\cite{cbR} and BES~\cite{besR}.
            The light shaded (green) band indicates the HVPTools average within 
            $1\,\sigma$ errors. 
}
\label{fig:xseccharm}
\end{figure}
The KLOE Collaboration has published new $\pp\gamma$ cross section data with \pp invariant 
mass-squared between 0.1 and $0.85\:\gev^2$~\cite{kloe10}. The radiative photon in 
this analysis is required to be detected in the electromagnetic calorimeter, 
which reduces the selected data sample to events with large photon scattering angle
(polar angle between $50^\circ$ and $130^\circ$), and photon energies above $20\:\mev$. 
The new data are found to be in agreement with, but less precise than, previously 
published data using small angle photon scattering~\cite{kloe08} (superseding earlier 
KLOE data~\cite{kloe04}). They hence exhibit the known discrepancy, on the $\rho$ 
resonance peak and above, with other \pp data, in particular those from BABAR, obtained 
using the same ISR technique~\cite{babarpipi}, and with data from $\taum\to\pipiz\nut$ 
decays~\cite{eetaunew}. 

Figure~\ref{fig:xsecpipi} shows the available $\ee\to\pp$ cross section measurements 
in various panels for different centre-of-mass energies ($\sqrt{s}$). The 
light shaded (green) band indicates the HVPTools average within $1\,\sigma$ errors.
The deviation between the average and the most precise individual measurements is 
depicted in Fig.~\ref{fig:comppipi}. Figure~\ref{fig:weights} shows the weights 
versus $\sqrt{s}$ the different experiments obtain in the locally performed average. 
BABAR and KLOE dominate the average over the entire energy range. 
Owing to the sharp radiator function, the available statistics for KLOE increases 
towards the $\phi$ mass, hence outperforming BABAR above $\sim$$0.8\:\gev$. For example, 
at $0.9\:\gev$ KLOE's small photon scattering angle data~\cite{kloe08} have statistical 
errors of $0.5\%$, which is twice smaller than that of BABAR (renormalising
BABAR to the 2.75 times larger KLOE bins at that energy). Conversely, at $0.6\:\gev$ the 
comparison reads $1.2\%$ (KLOE) versus $0.5\%$ (BABAR, again given in KLOE bins which 
are about 4.2 times larger than for BABAR at that energy). The discrepancy between the 
BABAR and KLOE data sets above $0.8\:\gev$ causes error rescaling in their 
average, and hence loss of precision. The group of experiments labelled ``other exp'' 
in Fig.~\ref{fig:weights} corresponds to older data with incomplete radiative 
corrections. Their weights are small throughout the entire energy domain. The 
computation of the dispersion integral over the full \pp spectrum requires 
to extend the available data to the region between threshold and $0.3\:\gev$, for 
which we use a fit as described in Ref.~\cite{g209}.

We have modified in this work the treatment of the $\omega(782)$ and $\phi(1020)$ 
resonances, using non-resonant data from BABAR~\cite{babar3pi}. 
While in our earlier analyses, the resonances
were fitted, analytically integrated, and the non-resonant contributions added 
separately, we now determine all the dominant contributions directly from the 
corresponding measurements. Hence the $\omega$ and $\phi$ contributions are included
in the $\pp\piz$, $\piz\gamma$, $\eta\gamma$, $\Kp\Km$, $\KS\KL$ spectra. Small 
remaining decay modes are considered separately. As an example for this procedure, 
the $\ee\to\pp\piz$ cross section measurements, featuring dominantly the $\omega$ 
and $\phi$ resonances,  are shown in Fig.~\ref{fig:xsec3pi}, together with the 
HVPTools average.

We also include new, preliminary, $\ee\to\pp 2\piz$ cross section measurements
from BABAR~\cite{babar2pi2pi0}, which significantly help
to constrain a contribution with disparate experimental information. The available 
four-pion measurements and HVPTools averages are depicted in Fig.~\ref{fig:xsec4pi} 
in linear (top) and logarithmic (bottom) ordinates.\footnote
{
\label{ftn:bcvc4pion}
   The new measurements also improve the conserved vector current (CVC) predictions 
   for the corresponding $\tau$ decays with four pions in the final state. 
   Coarse isospin-breaking corrections with 100\% uncertainty are 
   applied~\cite{4pioncorr}. We find 
   $\BR_{\rm CVC}(\taum\to\pim3\piz\nut)=(1.07\pm0.06)\%$, to be compared to 
   the world average of the direct measurements $(1.04\pm0.07)\%$~\cite{pdg10}, 
   and $\BR_{\rm CVC}(\taum\to2\pim\pip\piz\nut)=(3.79\pm0.21)\%$, to be compared to 
   the direct measurement $(4.48\pm0.06)\%$. The deviation between prediction 
   and measurement in the latter channel amounts to $3.2\,\sigma$, compared to 
   $3.6\,\sigma$ without the BABAR data~\cite{dehz02}. It is due to a 
   discrepancy in mainly the normalisation of the corresponding
   $\tau$ and \ee spectral functions. It is therefore important that the BABAR 
   and Belle experiments also perform these $\tau$ branching fraction 
   measurements as independent cross checks.
} 

The precise BABAR data~\cite{babar4pipi0,babar6pi,babarkkpi,babarkkpipi}
available for several higher multiplicity modes with and 
without kaons (which greatly benefit from the excellent particle identification
capabilities of the BABAR detector) help to discriminate between older, less precise
and sometimes contradicting measurements. Figure~\ref{fig:xsecmulti} shows the 
cross section measurements and HVPTools averages for the channels $\Kp\Km\pp$ 
(upper left), $2\pip2\pim\piz$ (upper right), $3\pip3\pim$ (lower left), 
and $2\pip2\pim2\piz$ 
(lower right). The BABAR data supercede much less precise measurements 
from M3N, DM1 and DM2. In several occurrences, these older measurements 
overestimate the cross sections in comparison with BABAR, which contributes to 
the reduction in the present evaluation of the hadronic loop effects. 

Finally, Fig.~\ref{fig:xseccharm} shows the charm resonance region above the 
opening of the $D\Dbar$ channel. Good agreement between the measurements is 
observed within the given errors. While Crystal Ball~\cite{cbR} and BES~\cite{besR}
published bare inclusive cross section results, PLUTO applied only radiative 
corrections~\cite{plutoradcorr}
following the formalism of Ref.~\cite{bonnmartin}, which does not include hadronic
vacuum polarisation. As in previous cases~\cite{dehz02} for the treatment of missing
radiative corrections in older data, we have applied this correction and assigned
a 50\% systematic error to it. 

\section{Missing hadronic channels}

Several five and six-pion modes involving \piz's, as well as $K\Kbar [n\pi]$ 
final states are still unmeasured. Owing to isospin invariance, their contributions
can be related to those of known channels. The new BABAR cross section data and 
results on process dynamics thereby allow more stringent constraints of 
the unknown contributions than the ones obtained in our previous 
analyses~\cite{dehz02,dehz03}.

\vspace{0.3cm}
\paragraph*{\bf\em Pais Isospin Classes.} 

Pais introduced~\cite{pais} a classification of $N$-pion states with total 
isospin $I=0,1$. The basis of isospin wave functions of a given state belongs to
irreducible representations of the corresponding symmetry group, which are 
characterised by three integer quantum numbers ({\em partitions} of $N$) 
$N_1,N_2,N_3$, obeying the relations $N_1\geq N_2\geq N_3\geq 0$ and $N_1+N_2+N_3=N$. 
The total isospin $I$ is determined uniquely to be $I=0$ if $N_1-N_2$ and $N_2-N_3$
are both even, and $I=1$ in the other cases. States $\{N_1,N_2,N_3\}$ are composed
by $N_3$ isoscalar three-pion subsystems, $N_2-N_3$ isovector two-pion subsystems,
and $N_1-N_2$ isovector single pions.

Simple examples are $\{ 110\}$ for $\ee\to\pp$ ($I=1$, $\rho$-like), and $\{ 111\}$ 
for $\ee\to\pp\piz$ ($I=0$, $\omega$-like). For four pions there are two channels 
and two isospin classes, related at the cross section level by\footnote
{
   In the following, and if not otherwise stated, $\sigma(X)$ denotes $\sigma(\ee\to X)$.
}
\beqn
   \sigma(\ee\to 2\pip2\pim)  &=& \frac{4}{5}\sigma_{310}\,, \\
   \sigma(\ee\to \pp 2\piz)   &=& \frac{1}{5}\sigma_{310} + \sigma_{211}\,.
\eeqn
The two isospin classes correspond to the resonant final states $\rho \pi\pi$
for $\{ 310\}$ and $\omega \piz$ for $\{ 211\}$. 

The $I=1$ states produced in \ee are related to the vector part of 
specific $\tau$ decays by isospin symmetry (CVC).

\vspace{0.3cm}
\paragraph*{\bf\em Five-Pion Channels.}

There are two five-pion final states, $2\pi^+ 2\pi^- \piz$ and $\pi^+ \pi^- 3\piz$, of which 
only the first has been measured. There is only one isospin class $\{ 311\}$, 
corresponding to $\omega \pi\pi$ and obeying the relation
$\sigma(2\pi^+ 2\pi^- \piz)=2\sigma(\pi^+ \pi^- 3\piz)=\frac{2}{3}\sigma_{311}$.

BABAR has shown~\cite{babar4pipi0} that the first channel is indeed dominated 
by $\omega \pi\pi$, with some contribution from $\eta \pi\pi$ via the 
isospin-violating decay $\eta \to \pp\piz$. These $\eta$ contributions
must be subtracted and treated separately as they do not obey the Pais
classification rules. At larger masses there is evidence for a $\rho \pi\pi$ 
component, which should correspond to $I_{3\pi}=0$ contributions above the 
$\omega$. Isospin symmetry holds for this contribution. 

The estimation procedure for the unknown five-pion contribution is as follows:
$\sigma(2\pi^+ 2\pi^- \piz)_{\mbox{\ftn$\eta$-excl}}=\sigma(2\pi^+ 2\pi^- \piz)-\sigma(\eta \pp)\times \BR(\eta\to \pp\piz)$,
with $\BR(\eta\to \pp\piz)=0.2274\pm0.0028$~\cite{pdg10},
$\sigma(\pi^+ \pi^- 3\piz)_{\mbox{\ftn$\eta$-excl}}=\frac{1}{2}\sigma(2\pi^+ 2\pi^- \piz)_{\mbox{\ftn$\eta$-excl}}$, 
and $\sigma(\eta \pp)$ is considered separately.
There is no contribution from $\sigma(\eta 2\piz)$, and
the contribution of $\omega \pi\pi$ with non purely pionic $\omega$ decays 
is taken from $\frac{3}{2}\sigma(\omega \pp)\times \BR(\mbox{$\omega$-non-pionic})$ 
with $\BR(\mbox{$\omega$-non-pionic})=0.093\pm0.007$~\cite{pdg10}.

\vspace{0.3cm}
\paragraph*{\bf\em Six-Pion Channels.}

There are three channels and four isospin classes with the relations ($I=1$):
\beqn
  \label{eq:6pi1}
  \sigma(3\pi^+ 3\pi^-)       &=& \frac{24}{35}\sigma_{510}+\frac{3}{5}\sigma_{330}\,, \\
  \label{eq:6pi2}
  \sigma(2\pi^+ 2\pi^- 2\piz) &=& \frac{8}{35}\sigma_{510}+\frac{2}{5}\sigma_{411}
                                  +\frac{2}{5}\sigma_{330}+\sigma_{321}\,,~~~~~ \\
  \label{eq:6pi3}
  \sigma(\pi^+ \pi^- 4\piz)&=&\frac{3}{35}\sigma_{510}+\frac{3}{5}\sigma_{411}\,,
\eeqn
where the lowest-mass resonant states are $\rho 4\pi$ for $\{ 510\}$, $\omega 3\pi$ for 
$\{ 411\}$, $3\rho$ for $\{ 330\}$, and $\omega \rho \pi$ for $\{ 321\}$.

BABAR has measured~\cite{babar6pi} the cross sections~(\ref{eq:6pi1}) and (\ref{eq:6pi2}), 
and observed only one $\rho$ state per event in the fully charged mode, thus 
favouring $\{ 510\}$ over $\{ 330\}$ in~(\ref{eq:6pi1}). The $\ee\to2\pi^+ 2\pi^- 2\piz$
process is dominated by $\omega \pp\piz$ up to $2\:\gev$. A small $\eta$
contribution is also found, but only the cross section for $\eta\omega$ is 
given.

To estimate the cross section~(\ref{eq:6pi3}) the relative contributions 
of $\{ 321\}$ and $\{ 411\}$ need to be known, which can be constrained
from $\tau$ data. The corresponding isospin relations for the $\tau$ spectral 
functions are
\beqn
   v(\taum\to2\pi^+ 3\pi^-\piz\nut)  &=& \frac{16}{35}v_{510}+\frac{4}{5}v_{411} \\\nonumber
                                    && +\;\frac{1}{5}v_{330}+\frac{1}{2}v_{321}\,, \\
   v(\taum\to\pi^+ 2\pi^- 3\piz\nut) &=& \frac{10}{35}v_{510}+\frac{1}{5}v_{411} \\\nonumber 
                                   && +\;\frac{4}{5}v_{330}+\frac{1}{2}v_{321}\,, \\
   v(\taum\to\pi^- 5\piz\nut)        &=& \frac{9}{35}v_{510}\,.
\eeqn
The branching fractions of the first two modes have been measured by CLEO. 
As for the \ee final states, they
are dominated by $\omega$ production, $\omega \pi^+2\pi^-$ and 
$\omega \pi^-2\piz$, with branching fractions $(1.20\pm0.22)\cdot10^{-4}$
and $(1.4\pm0.5)\cdot10^{-4}$, respectively, to be compared to total branching fractions 
of $(1.40\pm0.29)\cdot10^{-4}$ and $(1.83\pm0.50)\cdot10^{-4}$ ($\eta$ subtracted). 
This yields the bound $v_{411}/v_{321}<0.42$. 
The limit for $\pp4\piz$ is looser than that quoted in Ref.~\cite{dehz02}, 
where the $\{411\}$ partition was assumed to be negligible.

The estimation procedure for the missing six-pion mode is as follows: 
$\sigma(2\pi^+ 2\pi^- 2\piz)_{\mbox{\ftn$\eta$-excl}}=\sigma(2\pi^+ 2\pi^- 2\piz)-\sigma(\eta \omega)\times \BR(\eta\to \pp\piz)\times \BR(\omega \to \pp\piz)$, 
with $\BR(\omega\to \pp\piz)=0.892\pm0.007$~\cite{pdg10}, and
$\sigma(\pp4\piz)=0.0625\sigma(3\pi^+3\pi^-)+0.145\sigma(2\pi^+ 2\pi^- 2\piz)_{\mbox{\ftn$\eta$-excl} }~~~~\pm~100\%$;
$\sigma(\eta \pp\piz)=\sigma(\eta \omega)\times \BR(\omega \to \pp\piz)$ 
is treated separately, and the contribution from non-pionic $\omega$ decays is given by
$(1.145\pm0.145)\times\sigma(2\pi^+2\pi^-2\piz)\times \BR(\mbox{$\omega$-non-pionic})/\BR(\omega \to \pp\piz)$.

\vspace{0.3cm}
\paragraph*{\bf\em \boldmath$K\Kbar\pi$ Channels.}

The measured final states are $\KS K^\pm\pi^\mp$ and $K^+K^-\piz$, with 
$\KS\KL\piz$ missing ($C_{K\Kbar}=-1$). Except for a very small $\phi\piz$ 
contribution, these processes are governed by $K^0K^{\star0}(890)$ (dominant) and 
$K^\pm K^{\star\mp}(890)$ transitions below $2\:\gev$. Both $I=0,1$ amplitudes ($A_{0,1}$)
contribute. The fit of the Dalitz plot in the first channel yields the moduli 
of the two amplitudes and their relative phase as a function of mass. Hence
everything is determined, as seen from the following relations (labels written 
in the order $KK^\star$ with the given $K^\star$ decay modes):
\beqn
  \sigma(K^+K^-\piz+K^-K^+\piz)     &=& \frac{1}{6}|A_0-A_1|^2\,, \\
  \sigma(\KS\KL\piz+\KL\KS\piz)    &=& \frac{1}{6}|A_0+A_1|^2\,, \\
  \sigma(K^0K^-\pi^++\Kzb K^+\pi^-) &=& \frac{1}{3}|A_0+A_1|^2\,, \\
  \sigma(K^+\Kzb\pi^-+K^-K^0\pi^+)  &=& \frac{1}{3}|A_0-A_1|^2\,.
\eeqn 
The measured $\KS K^\pm\pi^\mp$ cross section (no ordering here) is therefore 
equal to $\frac{1}{3}[|A_0|^2+|A_1|^2]= \frac{1}{3}(\sigma_0+\sigma_1)$, and
$\sigma(K\Kbar\pi)=3\sigma(\KS K^\pm\pi^\mp)$ for the dominant $KK^\star$
part. Note that, unlike it was assumed in Ref.~\cite{dehz02,dehz03}, in general 
$\sigma(\KS\KL\piz)$ is not equal to $\sigma(K^+K^-\piz)$.

The complete $K\Kbar\pi$ contribution is obtained from
$\sigma(K\Kbar\pi)=3\sigma(\KS K^\pm\pi^\mp)+\sigma(\phi \piz)\times \BR(\phi\to K\Kbar)$, 
with $\BR(\phi\to K\Kbar)=0.831\pm0.003$, 
where contributions from non-hadronic $\phi$ decays are neglected, 
whereas decays to $\pp\piz$ are already counted in the multi-pion channels. 

\vspace{0.3cm}
\paragraph*{\bf\em \boldmath$K\Kbar 2\pi$ Channels.}

The channels measured by BABAR are $K^+K^-\pp$ and $K^+K^-2\piz$~\cite{babarkkpipi}.
They are dominated by $K^\star K\pi$, with $K\pi$ not in a 
$K^\star$, and smaller contributions from $K^+K^-\rho^0$ and $\phi \pi\pi$.

In the dominant $K^\star K\pi$ mode one can have $I=0$ and $I=1$ amplitudes. The different
charge configurations can be obtained via $I_{K\pi}=1/2$ and $3/2$ amplitudes, 
where, however, $I_{K\pi}=3/2$ is not favoured because it would have predicted
$\sigma(K^+K^-\pp)=\sigma(K^+K^-2\piz)$, whereas a ratio of roughly 4:1 has been 
measured~\cite{babarkkpipi}. In the following we assume a pure $I_{K\pi}=1/2$ state,
so that the relevant cross sections read (labels in the order $K^\star K\pi$,
appropriately summing over $K^0(\Kzb)$)
\beqn
  \sigma(K^\pm\piz K^\mp\piz)      &=& \frac{1}{18}|A_0-A_1|^2\,, \\
  \sigma(K^0\pi^\pm K^\mp\piz)     &=& \frac{1}{9}|A_0-A_1|^2\,, \\
  \sigma(K^0\piz K^0\piz)         &=& \frac{1}{18}|A_0+A_1|^2\,, \\
  \sigma(K^\pm\pi^\mp K^0\piz)     &=& \frac{1}{9}|A_0+A_1|^2\,, \\
  \sigma(K^0\piz K^\pm\pi^\mp)     &=& \frac{1}{9}|A_0+A_1|^2\,, \\
  \sigma(K^\pm\pi^\mp K^\pm\pi^\mp) &=& \frac{2}{9}|A_0+A_1|^2\,, \\
  \sigma(K^\pm\piz K^0\pi^\mp)     &=& \frac{1}{9}|A_0-A_1|^2\,, \\
  \sigma(K^0\pi^\pm K^0\pi^\mp)    &=& \frac{2}{9}|A_0-A_1|^2 \,.
\eeqn
This leads to $\sigma(K\Kbar\pi\pi)=9\sigma(K^+K^-\piz\piz)+\frac{9}{4}\sigma(K^+K^-\pp)$.

The inclusive $\sigma(K\Kbar\rho)$ cross section is thus obtained
as follows: get $\sigma(\phi \pp)=2\sigma(\phi 2\piz)$ and 
$\sigma(K^+K^-\rho^0)=\sigma(K^+K^-\pp)-\sigma(K^{\star0}K^\pm\pi^\mp)-\sigma(\phi \pp)\times \BR(\phi\to K^+K^-)$
(note that the published BABAR cross section table for $K^{\star0}K^\pm\pi^\mp$ already 
includes the branching fraction for $K^{\star0}\to K^\pm\pi^\mp$).
In lack of more information, we assume $\sigma(K\Kbar\rho)=4\sigma(K^+K^-\rho^0)$, 
with a 100\% error, and obtain
$\sigma(K\Kbar\pi\pi)=9[\sigma(K^+K^-2\piz)-\sigma(\phi 2\piz)]+\frac{9}{4}\sigma(K^{\star0}K^\pm\pi^\mp)+\frac{3}{2}\sigma(\phi\pp)+4\sigma(K^+K^-\rho^0)$.

\vspace{0.3cm}
\paragraph*{\bf\em \boldmath$K\Kbar 3\pi$ Channels.}

BABAR has only measured the final state $K^+K^-\pp\piz$~\cite{babar4pipi0}, which is 
dominated by $K^+K^-\omega$ up to $2\:\gev$. The channel $\phi\eta$ has been measured, 
and the remaining $\phi\pp\piz$ amplitude is negligible. The $\omega$
dominance does not apply to the missing channels $K^0K^\pm\pi^\mp\pp$
and $K^0K^\pm\pi^\mp 2\piz$, but their dynamics (for instance $K^\star$) should
be seen in the measured $K^+K^-\pp\piz$ mode, so it may be small, at
least below $2\:\gev$.

The missing channels are estimated as follows: 
$\sigma(K^+K^-\pp\piz)_{\mbox{\ftn$\eta$-excl}}=\sigma(K^+K^-\pp\piz)-\sigma(\phi\eta)\times \BR(\phi\to K^+K^-)\times \BR(\eta\to \pp\piz)$.
We assume, within a systematic error of 50\%, 
$\sigma(K^0\Kzb\pp\piz)_{\mbox{\ftn$\eta$-excl}}=\sigma(K^+K^-\pp\piz)_{\mbox{\ftn$\eta$-excl}}$, 
treat $\sigma(\phi\eta)$ separately, and compute the non-pionic $\omega$ contribution by 
$2\sigma(K^+K^-\pp\piz)_{\mbox{\ftn$\eta$-excl}}\times \BR(\mbox{$\omega$-non-pionic})/\BR(\omega \to \pp\piz)$.
Contributions from $K^0K^\pm\pi^\mp\pp$ and $K^0K^\pm\pi^\mp 2\piz$  
below $2\:\gev$ are neglected.

\vspace{0.3cm}
\paragraph*{\bf\em \boldmath$\eta 4\pi$ Channels.}

BABAR has measured $\sigma(\eta 2\pi^+2\pi^-)$~\cite{babar4pipi0}, where 
the $4\pi$ state has $C=-1$,
$I=1$. Because $\sigma(2\pi^+2\pi^-)\approx\sigma(\pp2\piz)$, we assume the
same ratio for the $\eta 4\pi$ process with the same $4\pi$ quantum numbers. 
We thus estimate $\sigma(\eta 4\pi)=2\sigma(\eta 2\pi^+2\pi^-)$, and 
assign a systematic error of 25\% to it.

\section{Data averaging and integration}

In this work, we have extended the use of HVPTools\footnote
{
   See Ref.~\cite{g209} for a more detailed description of the averaging and
   integration procedure developed for HVPTools. 
} 
to all experimental cross section data used in the compilation.\footnote
{
   So far~\cite{g209}, only the two-pion and four-pion channels were fully 
   evaluated using HVPTools, while all other contributions were taken from our previous 
   publications, using less sophisticated averaging software. 
}
The main difference of HVPTools 
with respect to our earlier software is that it replaces linear interpolation between 
adjacent data points (``trapezoidal rule'') by quadratic interpolation, which is 
found from toy-model analyses, with known truth integrals, to be more accurate. 
The interpolation functions are locally averaged between experiments, whereby 
correlations between measurement points of the same experiment and among different 
experiments due to common systematic errors are fully taken into account. Incompatible 
measurements lead to error rescaling in the local averages, using the PDG 
prescription~\cite{pdg10}. 

The errors in the average and in the integration for each channel are obtained 
from large samples of pseudo Monte Carlo experiments, by fluctuating all data points 
within errors and along their correlations. The integrals of the exclusive channels 
are then summed up, and the error of the sum is obtained by adding quadratically 
(linearly) all uncorrelated (correlated) errors.

Common sources of systematic errors also occur between measurements of different 
final state channels and must be taken into account when summing up the exclusive 
contributions. Such correlations mostly arise from luminosity uncertainties, if 
the data stem from the same experimental facility, and from radiative corrections. 
In total eight categories of correlated systematic uncertainties are distinguished.
Among those the most significant belong to radiative corrections, which are the same 
for CMD2 and SND, as well as to luminosity determinations by BABAR, CMD2 and SND 
(correlated per experiment for different channels, but independent between different 
experiments). 

\section{Results}

\begin{table}[t]
 \caption[.]{\label{tab:resultspipi}
   Contributions to \amuhadLO (middle column) from the individual \pp cross section 
   measurements by BABAR~\cite{babarpipi}, KLOE~\cite{kloe10,kloe08}, 
   CMD2~\cite{cmd203,cmd2new}, and SND~\cite{snd}. Also given are the corresponding CVC predictions
   of the $\tau\to\pim\piz\nut$ branching fraction (right column), corrected for isospin-breaking
   effects~\cite{eetaunew}. Here the first error is experimental and the second estimates the 
   uncertainty in the isospin-breaking corrections. The predictions are to be compared 
   with the world average of the 
   direct branching fraction measurements $(25.51 \pm 0.09)\%$~\cite{pdg10}.
   For each experiment, all available data in the energy range from threshold 
   to $1.8\:\gev$ ($m_\tau$ for $\BR_{\rm CVC}$) are used, and the missing part is 
   completed by the combined \ee data. The corresponding (integrand dependent) 
   fractions of the full integrals provided by a given experiment are given in 
   parentheses.
}
\setlength{\tabcolsep}{0.0pc}
\begin{tabularx}{\columnwidth}{@{\extracolsep{\fill}}lrr} 
\hline\noalign{\smallskip}
Experiment & \mc{1}{c}{$\amuhadLO~[10^{-10}]$} & \mc{1}{c}{$\BR_{\rm CVC}~[\%]$} \\
\noalign{\smallskip}\hline\noalign{\smallskip}
BABAR & $514.1\pm 3.8~ (1.00)$  & $ 25.15 \pm 0.18 \pm 0.22~ (1.00)$ \\
KLOE  & $503.1\pm 7.1~ (0.97)$  & $ 24.56 \pm 0.26 \pm 0.22~ (0.92)$   \\
CMD2  & $506.6\pm 3.9~ (0.89)$  & $ 24.96 \pm 0.21 \pm 0.22~ (0.96)$   \\
SND   & $505.1\pm 6.7~ (0.94)$  & $ 24.82 \pm 0.30 \pm 0.22~ (0.91)$  \\
\noalign{\smallskip}\hline
\end{tabularx}
\end{table}

\input compil_table

A compilation of all contributions to \amuhadLO and to \dahadZ, as well as the total 
results, are given in Table~\ref{tab:results}. The experimental errors are separated
into statistical, channel-specific systematic, and common systematic contributions
that are correlated with at least one other channel. 

Table~\ref{tab:resultspipi} quotes the specific contributions of the various 
$\ee\to\pp$ cross section measurements to \amuhadLO. Also given are the corresponding 
CVC-based $\tau\to\pim\piz\nut$ branching fraction predictions. The largest (smallest)
discrepancy of $2.7\,\sigma$ ($1.2\,\sigma$) between prediction and direct measurement 
is exhibited by KLOE (BABAR). It is interesting to note that the four $\amuhadLO[\pp]$ 
determinations in Table~\ref{tab:resultspipi} agree within errors (the overall $\chi^2$ of 
their average amounts to 3.2 for 3 degrees of freedom), whereas significant discrepancies 
are observed in the corresponding spectral functions~\cite{g209}. Since we cannot think 
of good reasons why systematic effects affecting the spectral functions should necessarily 
cancel in the integrals, we refrain from averaging the four values with a resulting smaller
error. The combined contribution is instead computed from local averages of the 
spectral function data that are subjected to local error rescaling in case of incompatibilities.

The contributions of the $J/\psi$ and $\psi(2S)$ resonances in 
Table~\ref{tab:results} are obtained by numerically integrating the corresponding 
undressed\footnote
{
   The undressing uses the BABAR programme {\em Afkvac} correcting for both leptonic 
   and hadronic vacuum polarisation effects. The correction factors amount to 
   $(1-\Pi(s))^2=0.956$ and $0.957$ for the $J/\psi$ and $\psi(2S)$, respectively. 
} 
Breit-Wigner lineshapes. Using instead the narrow-width approximation, 
$\sigma_R = 12\pi^2\Gamma_{ee}^0/M_R\cdot\delta(s-M_R^2)$, gives compatible results. 
The errors in the integrals are dominated by the knowledge of the corresponding
bare electronic width $\Gamma_{R\to ee}^0$.

\begin{figure}[t]
\begin{center}
\includegraphics[width=\figsize]{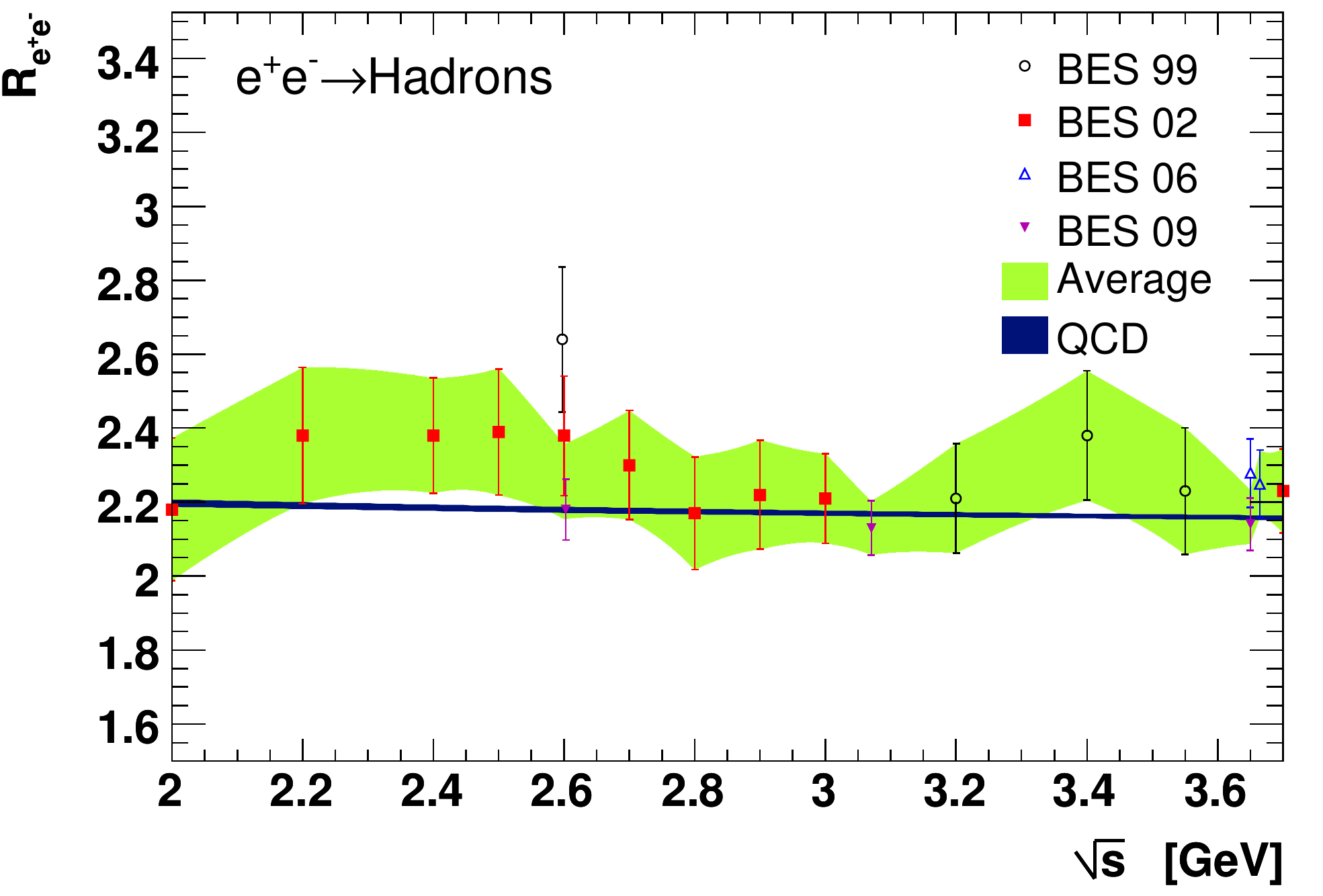}
\end{center}
\vspace{-0.5cm}
\caption[.]{ 
            Inclusive hadronic cross section ratio versus centre-of-mass energy in the 
            continuum region below the $D\Dbar$ threshold. Shown are bare
            BES data points~\cite{besR}, with statistical and systematic errors
            added in quadrature, the data average (shaded band), and the prediction 
            from massive perturbative QCD (solid line---see text). 
}
\label{fig:R}
\end{figure}

Sufficiently far from the quark thresholds we use four-loop~\cite{chetkuehn} 
perturbative QCD, including ${\cal O}(\as^2)$ quark mass corrections~\cite{kuhnmass}, 
to compute the inclusive hadronic cross section. Non-perturbative contributions at 
$1.8\:\gev$ were determined from data~\cite{dh98} and found to be 
small. The errors of the $R_{\rm QCD}$ contributions given in Table~\ref{tab:results}
account for the uncertainty in $\as$  (we use $\asZ=0.1193\pm0.0028$ from the fit
to the $Z$ hadronic width~\cite{gfitter}), the truncation of the perturbative 
series (we use the full four-loop contribution as systematic error), the full 
difference between fixed-order perturbation theory (FOPT) and, so-called, 
contour-improved perturbation theory (CIPT)~\cite{ledibpich}, as
well as quark mass uncertainties (we use the values and errors from Ref.~\cite{pdg10}).
The former three errors are taken to be fully correlated between the 
various energy regions (see Table~\ref{tab:results}), 
whereas the (smaller) quark-mass uncertainties
are taken to be uncorrelated. Figure~\ref{fig:R} shows the comparison 
between BES data~\cite{besR} and the QCD prediction below the $D\Dbar$ 
threshold between 2 and $3.7\:\gev$. Agreement within errors is found.\footnote
{
   To study the transition region between the sum of exclusive measurements and 
   QCD, we have computed \amuhadLO in two narrow energy intervals around 1.8\:\gev.
   For the energy interval $1.75$--$1.8\:\gev$ we find (in units of $10^{-10}$)
   $2.74 \pm 0.06 \pm 0.21$ (statistical and systematic errors) for the sum of 
   the exclusive data, and $2.53 \pm 0.03$ for perturbative QCD (see text for 
   the contributions to the error). For the interval $1.8$--$2.0\:\gev$ we find 
   $8.28  \pm 0.11 \pm 0.74$ and $8.31 \pm 0.09$ for data and QCD, respectively. 
   The excellent agreement represents another support for the use of QCD beyond 
   1.8\:\gev centre-of-mass energy. 
   Comparing the \amuhadLO predictions in the energy interval $2$--$3.7\:\gev$, 
   we find $26.5 \pm 0.2 \pm 1.7$ for BES data, and $25.2 \pm 0.2$ for perturbative 
   QCD. 
}

\vspace{0.3cm}
\paragraph*{\bf\em Muon magnetic anomaly.}

Adding all lowest-order hadronic contributions together yields the estimate
(this and all following numbers in this and the next paragraph are in units 
of $10^{-10}$)
\beq
\label{eq:amuhadlo}
   \amuhadLO = 692.3 \pm 4.2
\eeq
which is dominated by experimental systematic uncertainties (\cf Table~\ref{tab:results}
for a separation of the error into subcomponents). The new result is $-3.2\cdot 10^{-10}$ 
below that of our previous evaluation~\cite{g209}. This shift 
is composed of
$-0.7$ from the inclusion of the new, large photon angle data from KLOE, 
$+0.4$ from the use of preliminary BABAR data in the $\ee\to\pp 2\piz$ mode, 
$-2.4$ from the new high-multiplicity exclusive channels, the reestimate of 
the unknown channels, and the new resonance treatment, $-0.5$ from mainly the 
four-loop term in the QCD prediction of the hadronic cross section that 
contributes with a negative sign, as well as smaller other differences. 
The total error on \amuhadLO is slightly larger than that of Ref.~\cite{g209}
owing to a more thorough (and conservative) evaluation of the inter-channel
correlations.

\begin{figure}[t]
\includegraphics[width=\columnwidth]{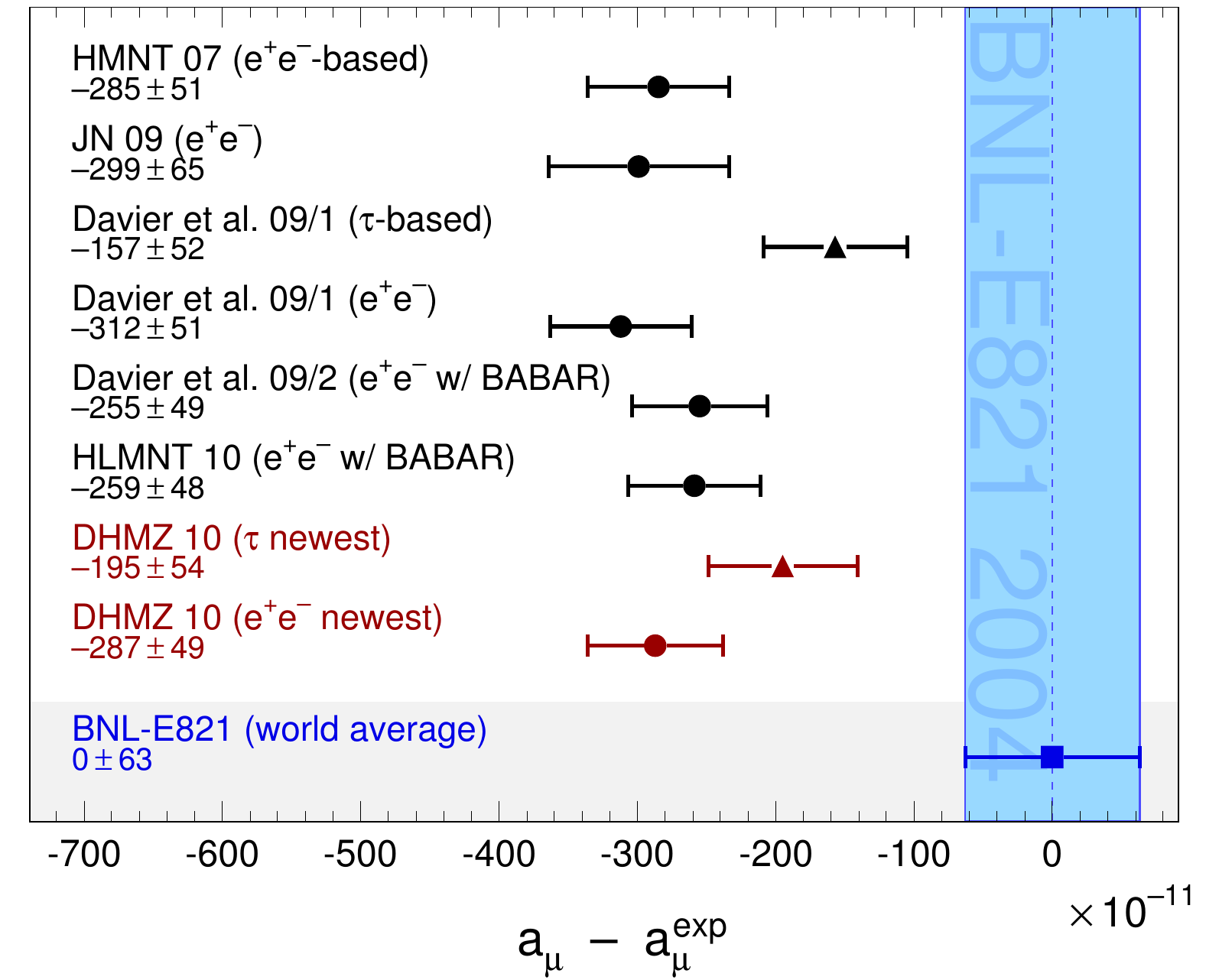}
\vspace{-0.5cm}
\caption{ 
        Compilation of recent results for $\amuSM$ (in units of $10^{-11}$),
        subtracted by the central value of the experimental average~\cite{bnl,pdgg-2rev}.
        The shaded vertical band indicates the experimental error. 
        The SM predictions are taken from: 
        this work (DHMZ 10), 
        HLMNT~\cite{hlmnt} (\ee based, including BABAR and KLOE 2010 \pp data),
        Davier \ea 09/1~\cite{eetaunew} (\Tau-based), 
        Davier \ea 09/1~\cite{eetaunew} (\ee-based, not including BABAR \pp data),
        Davier \ea 09/2~\cite{g209} (\ee-based including BABAR \pp data),
        HMNT 07~\cite{hmnt} and JN 09~\cite{jeger} (not including BABAR \pp data).}
\label{fig:amures}
\end{figure}
Adding to the result~(\ref{eq:amuhadlo}) the contributions from higher order 
hadronic loops, $-9.79 \pm 0.09$~\cite{hmnt}, hadronic light-by-light 
scattering, $10.5\pm 2.6$~\cite{prades09} (\cf remark in Footnote~\ref{ftn:lbls}),
as well as QED, $11\,658\,471.809 \pm 0.015$~\cite{kinoshita} (see also~\cite{pdgg-2rev} 
and references therein), and electroweak effects,
$15.4 \pm 0.1_{\rm had} \pm 0.2_{\rm Higgs}$~\cite{jackiw,czarnecki,knecht},
we obtain the SM prediction
\beq
\label{eq:amusm}
  \amuSM = 11\,659\,180.2 \pm 4.2 \pm 2.6 \pm 0.2~(4.9_{\rm tot})\,,
\eeq
where the errors account for lowest and higher order hadronic, and 
other contributions, respectively. The result~(\ref{eq:amusm}) deviates from the 
experimental average, $\amuExp=11\,659\,208.9 \pm 5.4 \pm 3.3$~\cite{bnl,pdgg-2rev}, 
by $28.7 \pm 8.0$ ($3.6\,\sigma$).\footnote
{\label{ftn:lbls}Using the alternative result for the light-by-light scattering 
   contribution, $11.6\pm 4.0$~\cite{nyffeler}, the error in the SM 
   prediction~(\ref{eq:amusm}) increases to $5.8$, and the discrepancy with 
   experiment reduces to $3.2\sigma$.
} 

A compilation of recent SM predictions for \amu compared with the experimental
result is given in Fig.~\ref{fig:amures}.

\vspace{0.3cm}
\paragraph*{\bf\em Update of \boldmath$\tau$-based $g-2$ result.}

The majority of the changes applied in this work, compared to our previous one~\cite{g209}, 
will similarly affect the $\tau$-based result from Ref.~\cite{eetaunew}, requiring
a reevaluation of the corresponding $\tau$-based hadronic contribution. 
In the $\tau$-based analysis~\cite{adh}, the \pp cross section is entirely replaced by 
the average, isospin-transformed, and isospin-breaking corrected 
$\tau\to\pim\piz\nut$ spectral function,\footnote
{
   Using published $\tau\to\pim\piz\nut$ spectral function data from 
   ALEPH~\cite{taualeph}, Belle~\cite{taubelle}, 
   CLEO~\cite{taucleo} and OPAL~\cite{tauopal}, and using the world 
   average branching fraction~\cite{pdg10} (2009 PDG edition).
}
while the four-pion cross sections, obtained from linear combinations of the 
$\taum\to\pim3\piz\nut$ and $\taum\to2\pim\pip\piz\nut$ spectral functions,\footnote
{
   Similar to Footnote~\ref{ftn:bcvc4pion}, 
   coarse isospin-breaking corrections with 100\% uncertainty are applied to 
   the four-pion spectral functions from $\tau$ decays~\cite{4pioncorr}. 
} 
are only evaluated up to $1.5\:\gev$ with $\tau$ data. Due to the lack of 
statistical precision, the spectrum is completed with \ee data between 1.5 and 1.8\:\gev. 
All the other channels are taken from \ee data. The complete lowest-order 
$\tau$-based result reads
\beq
\label{eq:amuhadlotau}
   \amuhadLO[\tau] = 701.5 \pm 3.5 \pm 1.9 \pm 2.4 \pm 0.2 \pm 0.3\,,
\eeq
where the first error is $\tau$ experimental, the second estimates the uncertainty 
in the isospin-breaking corrections, the third is \ee experimental, and 
the fourth and fifth stand for the narrow resonance and QCD uncertainties, respectively. 
The $\tau$-based hadronic contribution deviates by $9.1 \pm 5.0$ ($1.8\,\sigma$) from the 
\ee-based one, and the full $\tau$-based SM prediction $\amuSM[\tau]=11\,659\,189.4 \pm 5.4$
deviates by $19.5 \pm 8.3$ ($2.4\,\sigma$) from the experimental average. The new $\tau$-based 
result is also included in the compilation of Fig.~\ref{fig:amures}.

\vspace{0.3cm}
\paragraph*{\bf\em Running electromagnetic coupling at \boldmath$M_Z^2$.}

\begin{figure}[t]
\includegraphics[width=\columnwidth]{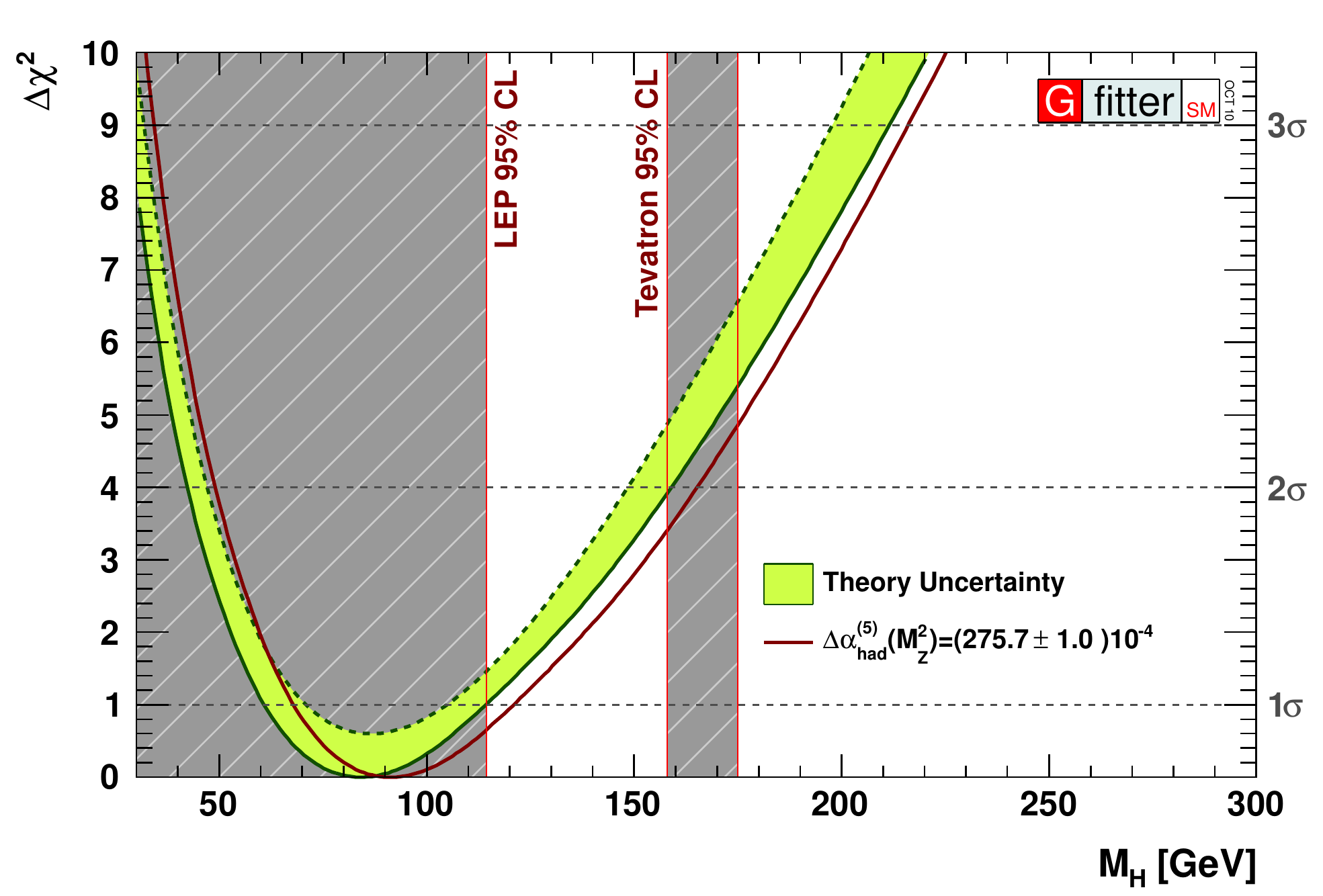}
\vspace{-0.5cm}
\caption{ 
         Standard Gfitter electroweak fit result~\cite{gfitter} (green 
         shaded band) and the result obtained for the new evaluation of 
         \dahadZ (red solid curve). The legend displays the corresponding five-quark 
         contribution, \dahadZf, where the top term of $-0.72\cdot10^{-4}$ is excluded. 
         A shift of $+7\:\gev$ in the central value of the Higgs boson
         is observed.
         }
\label{fig:gfitter}
\end{figure}
The sum of all hadronic contributions from Table~\ref{tab:results}
gives for the \ee-based hadronic term in the running of \aZ
\beq
\label{eq:dahad}
   \dahadZ   = (275.0 \pm 1.0)\cdot 10^{-4}\,,
\eeq
which is, contrary to the evaluation of \amuhadLO, not dominated by the 
uncertainty in the experimental low-energy data, but by contributions 
from all energy regions, where both experimental and theoretical errors 
have similar magnitude.\footnote
{
   In the global electroweak fit both $\as(M_Z)$ and \dahadZ are floating 
   parameters (though the latter one is constrained to its phenomenological 
   value). It is therefore important to include their mutual dependence in the 
   fit. The functional dependence of the central value of \dahadZ on the value 
   of \asZ approximately reads $0.37\cdot10^{-4}\times(\asZ-0.1193)/0.0028$.
} 
The corresponding $\tau$-based result reads 
$\dahadZ=(276.1 \pm 1.1)\cdot 10^{-4}$. As expected, the result~(\ref{eq:dahad}) 
is smaller than the most recent value from the HLMNT 
group~\cite{hlmnt} $\dahadZ=(275.5 \pm 1.4)\cdot 10^{-4}$. Owing to the use of 
perturbative QCD between 1.8 and $3.7\:\gev$, the precision in Eq.~(\ref{eq:dahad}) 
is  significantly improved compared to the HLMNT result, which relies 
on experimental data in that domain.\footnote
{
   HLMNT use perturbative QCD for the central value of the contribution 
   between 1.8 and $3.7\:\gev$, but assign the experimental errors from 
   the BES measurements to it. 
} 

Adding the three-loop leptonic contribution,
$\Delta\alpha_{\rm lep}(M_Z^2)=314.97686\cdot 10^{-4}$~\cite{steinhauser}, 
with negligible uncertainties, one finds
\beq
   \alpha^{-1}(M_Z^2) = 128.952 \pm 0.014\,.
\eeq

The running electromagnetic coupling at $M_Z$ enters at various levels the 
global SM fit to electroweak precision data. It contributes to the radiator 
functions that modify the vector and axial-vector couplings in the partial 
$Z$ boson widths to fermions, and also to the SM prediction of the 
$W$ mass and the effective weak mixing angle. Overall, the fit exhibits
a $-39\%$ correlation between the Higgs mass ($M_H$) and $\dahadZ$~\cite{gfitter}, 
so that the decrease in the value~(\ref{eq:dahad}) and thus in the running
electromagnetic coupling strength, with respect to earlier evaluations,  
leads to an increase in the most probable value 
of $M_H$ returned by the fit.\footnote
{
   The correlation between $M_H$ and $\dahadZ$ reduces to $-17\%$ when 
   using the result~(\ref{eq:dahad}) in the global fit. 
} 
Figure~\ref{fig:gfitter} shows the standard
Gfitter result (green shaded band)~\cite{gfitter}, using as hadronic  
contribution $\dahadZ=(276.8\pm2.2)\cdot 10^{-4}$~\cite{hmnt}, together with 
the result obtained by using Eq.~(\ref{eq:dahad}) (red solid line).
The fitted Higgs mass shifts from previously $84^{\,+30}_{\,-23}\:\gev$ 
to $91^{\,+30}_{\,-23}\:\gev$. The stationary error of the 
latter value, in spite of the improved accuracy in \dahadZ, is due to the 
logarithmic $M_H$ dependence of the fit observables. The new 95\% and 99\%
upper limits on $M_H$ are $163\:\gev$ and $193\:\gev$, respectively. 

\section{Conclusions}

We have updated the Standard Model predictions of the muon anomalous magnetic
moment and the running electromagnetic coupling constant at $M_Z^2$ by reevaluating
their virtual hadronic contributions. Mainly the reestimation of missing higher 
multiplicity channels, owing to new results from BABAR, causes a decrease 
of this contribution with respect to earlier calculations, which---on one 
hand---amplifies the discrepancy of the muon $g-2$ measurement with its prediction 
to $3.6\,\sigma$ for \ee-based analysis, and to $2.4\,\sigma$ for the $\tau$-based
analysis, while---on the other hand---relaxes the tension between the direct 
Higgs searches and the electroweak fit by $7\:\gev$ for the Higgs mass.

A thorough reestimation
of inter-channel correlations has led to a slight increase in the final error
of the hadronic contribution to the muon $g-2$. A better precision is 
currently constricted by the discrepancy between KLOE and the other experiments,
in particular BABAR, in the dominant \pp mode. This discrepancy is corroborated
when comparing \ee and $\tau$ data in this mode, where agreement between 
BABAR and $\tau$ data is observed. 

Support for the KLOE results must come from a cross-section measurement involving 
the ratio of pion-to-muon pairs. Moreover, new \pp precision data are soon expected 
from the upgraded VEPP-2000 storage ring at BINP-Novosibirsk, Russia, and the improved 
detectors CMD-3 and SND-2000. The future development of this field also relies 
on a more accurate muon $g-2$ measurement, and on progress in the evaluation of
the light-by-light scattering contribution. 

\begin{details}
We are most grateful to Martin Goebel from the Gfitter group for performing 
the electroweak fit and producing Fig.~\ref{fig:gfitter} of this paper. 
We thank Changzheng Yuan, Andreas B\"acker and Claus Grupen for information 
on the radiative corrections applied to BES and PLUTO data. This work is 
supported in part by the Talent Team Program of CAS (KJCX2-YW-N45) of China.
\end{details}

\vfill

\input Bibliography
\end{document}

%% file: defs.tex
\mathchardef\Upsilon="7107
\def\Y#1S{\ensuremath{\Upsilon{(#1S)}}\xspace}% no space before {...}!

\newcommand{\mZ}{\ensuremath{M_Z}\xspace}
\newcommand{\as}{\ensuremath{\alpha_{\scriptscriptstyle S}}\xspace}

\newcommand{\asZ}{\ensuremath{\as(\mZ^2)}\xspace}

\newcommand{\aZ}{\ensuremath{\alpha(M_Z^2)}\xspace}
\newcommand{\dahadZ}{\ensuremath{\Delta\alpha_{\rm had}(M_Z^2)}\xspace}
\newcommand{\dahadZf}{\ensuremath{\Delta\alpha_{\rm had}^{(5)}(M_Z^2)}\xspace}

\newcommand{\Kbar    }{\kern 0.2em\overline{\kern -0.2em K}{}\xspace}
\newcommand{\Dbar    }{\kern 0.2em\overline{\kern -0.2em D}{}\xspace}

\newcommand{\Kz      }{\ensuremath{K^0}\xspace}
\newcommand{\Kzb     }{\ensuremath{\Kbar^0}\xspace}
\newcommand{\KzKzb   }{\ensuremath{\Kz \kern -0.16em \Kzb}\xspace}
\newcommand{\Kp      }{\ensuremath{K^+}\xspace}
\newcommand{\Km      }{\ensuremath{K^-}\xspace}

\newcommand{\KpKm    }{\ensuremath{\Kp \kern -0.16em \Km}\xspace}
\newcommand{\KS      }{\ensuremath{K^0_{\scriptscriptstyle S}}\xspace} 
\newcommand{\KL      }{\ensuremath{K^0_{\scriptscriptstyle L}}\xspace}

\newcommand{\mc}{\multicolumn}

\newcommand{\Tau}{\ensuremath{\tau}\xspace}
\newcommand{\nut}{\ensuremath{\nu_\tau}\xspace}

\newcommand{\BR}{\ensuremath{{\cal B}}\xspace}

\newcommand{\pip}{\ensuremath{\pi^+}\xspace}
\newcommand{\pim}{\ensuremath{\pi^-}\xspace}
\newcommand{\piz}{\ensuremath{\pi^0}\xspace}

\newcommand{\ee}{\ensuremath{e^+e^-}\xspace}

\newcommand{\pp}{\ensuremath{\pi^+\pi^-}\xspace}
\newcommand{\pipiz}{\ensuremath{\pi^-\pi^0}\xspace}

\newcommand{\ppg}{\ensuremath{\pi^+\pi^-(\gamma)}\xspace}

\newcommand{\tev}{\ensuremath{\mathrm{\,Te\kern -0.1em V}}\xspace}
\newcommand{\gev}{\ensuremath{\mathrm{\,Ge\kern -0.1em V}}\xspace}
\newcommand{\mev}{\ensuremath{\mathrm{\,Me\kern -0.1em V}}\xspace}
\newcommand{\kev}{\ensuremath{\mathrm{\,ke\kern -0.1em V}}\xspace}
\newcommand{\ev}{\ensuremath{\mathrm{\,e\kern -0.1em V}}\xspace}
\newcommand{\gevc}{\ensuremath{{\mathrm{\,Ge\kern -0.1em V\!/}c}}\xspace}
\newcommand{\mevc}{\ensuremath{{\mathrm{\,Me\kern -0.1em V\!/}c}}\xspace}
\newcommand{\gevcc}{\ensuremath{{\mathrm{\,Ge\kern -0.1em V\!/}c^2}}\xspace}
\newcommand{\mevcc}{\ensuremath{{\mathrm{\,Me\kern -0.1em V\!/}c^2}}\xspace}

\newcommand{\bei}{\begin{itemize}}
\newcommand{\eei}{\end{itemize}}
\newcommand{\ben}{\begin{enumerate}}
\newcommand{\een}{\end{enumerate}}
\newcommand{\beq}{\begin{equation}}
\newcommand{\eeq}{\end{equation}}
\newcommand{\beqn}{\begin{eqnarray}}
\newcommand{\eeqn}{\end{eqnarray}}
\newcommand{\beqns}{\begin{eqnarray*}}
\newcommand{\eeqns}{\end{eqnarray*}}

\renewcommand{\arraystretch}{1.25}
\newcommand{\amu}{\ensuremath{a_\mu}\xspace}

\newcommand{\amuhadLO}{\ensuremath{\amu^{\rm had,LO}}\xspace}

\newcommand{\amuSM}{\ensuremath{\amu^{\rm SM}}\xspace}
\newcommand{\amuExp}{\ensuremath{\amu^{\rm exp}}\xspace}
\newcommand\cf{{cf.}\xspace}
\newcommand{\ea}{{\em et al.}\xspace}
\newcommand{\taum}{\tau^-}

\newcommand{\ftn}{\footnotesize}
\def\@citex[#1]#2{\if@filesw\immediate\write\@auxout{\string\citation{#2}}\fi
  \@tempcnta\z@\@tempcntb\m@ne\def\@citea{}\@cite{\@for\@citeb:=#2\do
    {\@ifundefined
       {b@\@citeb}{\@citeo\@tempcntb\m@ne\@citea
        \def\@citea{,\penalty\@m\ }{\bf ?}\@warning
       {Citation `\@citeb' on page \thepage \space undefined}}%
    {\setbox\z@\hbox{\global\@tempcntc0\csname b@\@citeb\endcsname\relax}%
     \ifnum\@tempcntc=\z@ \@citeo\@tempcntb\m@ne
       \@citea\def\@citea{,\penalty\@m}
       \hbox{\csname b@\@citeb\endcsname}%
     \else
      \advance\@tempcntb\@ne
      \ifnum\@tempcntb=\@tempcntc
      \else\advance\@tempcntb\m@ne\@citeo
      \@tempcnta\@tempcntc\@tempcntb\@tempcntc\fi\fi}}\@citeo}{#1}}

\def\@citeo{\ifnum\@tempcnta>\@tempcntb\else\@citea
  \def\@citea{,\penalty\@m}%
  \ifnum\@tempcnta=\@tempcntb\the\@tempcnta\else
   {\advance\@tempcnta\@ne\ifnum\@tempcnta=\@tempcntb \else
\def\@citea{--}\fi
    \advance\@tempcnta\m@ne\the\@tempcnta\@citea\the\@tempcntb}\fi\fi}
\newenvironment{myquote}
               {\list{}{\leftmargin0cm\indent}%
                \item\relax}
               {\endlist}
\newcommand\allFontSize{\footnotesize}
\newcommand\detailsSize{\allFontSize}
\newenvironment{details}%
{\begin{myquote}\detailsSize}{\end{myquote}}

%% file: compil_table.tex
\begin{table*}[p]
  \caption[.]{\label{tab:results}
    Compilation of the exclusive and inclusive contributions to \amuhadLO and \dahadZ.
    Where three (or more) errors are given, the first is statistical, the second
    channel-specific systematic, and the third common systematic, which 
    is correlated with at least one other channel. For the contributions 
    computed from QCD, only total errors are given, which include effects
    from the $\as$ uncertainty, the truncation of the perturbative series at four loops, 
    the FOPT vs.\ CIPT ambiguity (see text), and quark mass uncertainties. 
    Apart from the latter uncertainty, all other errors are taken to be 
    fully correlated among the various energy regions where QCD is used. 
    The errors in the Breit-Wigner 
    integrals of the narrow resonances $J/\psi$ and $\psi(2S)$ are dominated 
    by the uncertainties in their respective electronic width measurements~\cite{pdg10}. 
    The error on the sum (last line) is obtained by quadratically adding 
    all statistical and channel-specific systematic errors, and by linearly 
    adding correlated systematic errors where applies (see text for details 
    on the treatment of correlations between different channels). \\
}
\setlength{\tabcolsep}{0.0pc}
\begin{tabularx}{\textwidth}{@{\extracolsep{\fill}}lrr} 
\hline\noalign{\smallskip}
Channel &   \amuhadLO $[10^{-10}]$ & \dahadZ $[10^{-4}]$ \\
\noalign{\smallskip}\hline\noalign{\smallskip}
$\pi^0\gamma$                                          &$  4.42 \pm 0.08 \pm 0.13 \pm 0.12$&$  0.36 \pm 0.01 \pm 0.01 \pm 0.01$\\
$\eta\gamma$                                           &$  0.64 \pm 0.02 \pm 0.01 \pm 0.01$&$  0.08 \pm 0.00 \pm 0.00 \pm 0.00$\\
$\pi^+\pi^-$                                           &$507.80 \pm 1.22 \pm 2.50 \pm 0.56$&$ 34.43 \pm 0.07 \pm 0.17 \pm 0.04$\\
$\pi^+\pi^-\pi^0$                                      &$ 46.00 \pm 0.42 \pm 1.03 \pm 0.98$&$  4.58 \pm 0.04 \pm 0.11 \pm 0.09$\\
$2\pi^+2\pi^-$                                         &$ 13.35 \pm 0.10 \pm 0.43 \pm 0.29$&$  3.49 \pm 0.03 \pm 0.12 \pm 0.08$\\
$\pi^+\pi^-2\pi^0$                                     &$ 18.01 \pm 0.14 \pm 1.17 \pm 0.40$&$  4.43 \pm 0.03 \pm 0.29 \pm 0.10$\\
$2\pi^+2\pi^-\pi^0$ ($\eta$ excl.)                     &$  0.72 \pm 0.04 \pm 0.07 \pm 0.03$&$  0.22 \pm 0.01 \pm 0.02 \pm 0.01$\\
$\pi^+\pi^-3\pi^0$ ($\eta$ excl., from isospin)        &$  0.36 \pm 0.02 \pm 0.03 \pm 0.01$&$  0.11 \pm 0.01 \pm 0.01 \pm 0.00$\\
$3\pi^+3\pi^-$                                         &$  0.12 \pm 0.01 \pm 0.01 \pm 0.00$&$  0.04 \pm 0.00 \pm 0.00 \pm 0.00$\\
$2\pi^+2\pi^-2\pi^0$ ($\eta$ excl.)                    &$  0.70 \pm 0.05 \pm 0.04 \pm 0.09$&$  0.25 \pm 0.02 \pm 0.02 \pm 0.03$\\
$\pi^+\pi^-4\pi^0$ ($\eta$ excl., from isospin)        &$  0.11 \pm 0.01 \pm 0.11 \pm 0.00$&$  0.04 \pm 0.00 \pm 0.04 \pm 0.00$\\
$\eta\pi^+\pi^-$                                       &$  1.15 \pm 0.06 \pm 0.08 \pm 0.03$&$  0.33 \pm 0.02 \pm 0.02 \pm 0.01$\\
$\eta\omega$                                           &$  0.47 \pm 0.04 \pm 0.00 \pm 0.05$&$  0.15 \pm 0.01 \pm 0.00 \pm 0.02$\\
$\eta 2\pi^+2\pi^-$                                    &$  0.02 \pm 0.01 \pm 0.00 \pm 0.00$&$  0.01 \pm 0.00 \pm 0.00 \pm 0.00$\\
$\eta\pi^+\pi^-2\pi^0$ (estimated)                 &$  0.02 \pm 0.01 \pm 0.01 \pm 0.00$&$  0.01 \pm 0.00 \pm 0.00 \pm 0.00$\\
$\omega\pi^0~(\omega\to\pi^0\gamma)$           &$  0.89 \pm 0.02 \pm 0.06 \pm 0.02$&$  0.18 \pm 0.00 \pm 0.02 \pm 0.00$\\
$\omega\pi^+\pi^-,\omega 2\pi^0~(\omega\to\pi^0\gamma)$      &$  0.08 \pm 0.00 \pm 0.01 \pm 0.00$&$  0.03 \pm 0.00 \pm 0.00 \pm 0.00$\\
$\omega$ (non-$3\pi,\pi\gamma,\eta\gamma$)         &$  0.36 \pm 0.00 \pm 0.01 \pm 0.00$&$  0.03 \pm 0.00 \pm 0.00 \pm 0.00$\\
$K^+K^-$                                               &$ 21.63 \pm 0.27 \pm 0.58 \pm 0.36$&$  3.13 \pm 0.04 \pm 0.08 \pm 0.05$\\
$\KS\KL$                                               &$ 12.96 \pm 0.18 \pm 0.25 \pm 0.24$&$  1.75 \pm 0.02 \pm 0.03 \pm 0.03$\\
$\phi$ (non-$K\Kbar,3\pi,\pi\gamma,\eta\gamma$)&$  0.05 \pm 0.00 \pm 0.00 \pm 0.00$&$  0.01 \pm 0.00 \pm 0.00 \pm 0.00$\\
$K\Kbar\pi$ (partly from isospin)                &$  2.39 \pm 0.07 \pm 0.12 \pm 0.08$&$  0.76 \pm 0.02 \pm 0.04 \pm 0.02$\\
$K\Kbar2\pi$  (partly from isospin)             &$  1.35 \pm 0.09 \pm 0.38 \pm 0.03$&$  0.48 \pm 0.03 \pm 0.14 \pm 0.01$\\
$K\Kbar3\pi$  (partly from isospin)             &$ -0.03 \pm 0.01 \pm 0.02 \pm 0.00$&$ -0.01 \pm 0.00 \pm 0.01 \pm 0.00$\\
$\phi\eta$                                             &$  0.36 \pm 0.02 \pm 0.02 \pm 0.01$&$  0.13 \pm 0.01 \pm 0.01 \pm 0.00$\\
$\omega K\Kbar~(\omega\to\pi^0\gamma)$  &$  0.00 \pm 0.00 \pm 0.00 \pm 0.00$&$  0.00 \pm 0.00 \pm 0.00 \pm 0.00$\\
\noalign{\smallskip}\hline\noalign{\smallskip}
$J/\psi$ (Breit-Wigner integral)                                              &$  6.22 \pm 0.16$&$  7.03 \pm 0.18$\\
$\psi(2S)$ (Breit-Wigner integral)                                             &$  1.57 \pm 0.03$&$  2.50 \pm 0.04$\\
\noalign{\smallskip}\hline\noalign{\smallskip}
$R_{\rm data}~~[3.7-5.0\:\gev]$                       &$  7.29 \pm 0.05 \pm 0.30 \pm 0.00$&$ 15.79 \pm 0.12 \pm 0.66 \pm 0.00$\\
\noalign{\smallskip}\hline\noalign{\smallskip}
$R_{\rm QCD}~~[1.8-3.7\:\gev]_{uds}$                   &$ 33.45 \pm 0.28$&$ 24.27 \pm 0.19$     \\  
$R_{\rm QCD}~~[5.0-9.3\:\gev]_{udsc}$                  &$  6.86 \pm 0.04$&$ 34.89 \pm 0.18$      \\ 
$R_{\rm QCD}~~[9.3-12.0\:\gev]_{udscb}$                &$  1.21 \pm 0.01$&$ 15.56 \pm 0.04$       \\
$R_{\rm QCD}~~[12.0-40.0\:\gev]_{udscb}$               &$  1.64 \pm 0.01$&$ 77.94 \pm 0.12$       \\
$R_{\rm QCD}~~[>40.0\:\gev]_{udscb}$                   &$  0.16 \pm 0.00$&$ 42.70 \pm 0.06$       \\
$R_{\rm QCD}~~[>40.0\:\gev]_{t}$                       &$  0.00 \pm 0.00$&$ -0.72 \pm 0.01$       \\
\noalign{\smallskip}\hline\noalign{\smallskip}
{\bf Sum}                                         
      & $692.3 \pm 1.4 \pm 3.1 \pm 2.4 \pm 0.2_{\psi} \pm 0.3_{\rm QCD}$
      & $274.97 \pm 0.17  \pm 0.78 \pm 0.37 \pm 0.18_{\psi} \pm 0.52_{\rm QCD}$ \\
\noalign{\smallskip}\hline
\end{tabularx}
\end{table*}